\documentclass[9pt,twocolumn]{article}
\usepackage{newtxtext,newtxmath}
\usepackage{graphicx}
\usepackage{placeins}

\usepackage{physics}
\usepackage[letterpaper,margin=1in]{geometry}

\renewenvironment{abstract}
	{\quotation}
	{\endquotation}

\date{}

\makeatletter
\renewcommand{\fnum@figure}{\textbf{Figure \thefigure}}
\renewcommand{\fnum@table}{\textbf{Table \thetable}}
\makeatother

\usepackage{scicite}

\usepackage{url}

\definecolor{mycolor}{RGB}{219, 48, 122}

\def\scititle{Bicontinuity in active phase separation}
\title{\bfseries \boldmath \scititle}

\author{
Paarth Gulati$^{1,4\dagger}$,
Liang Zhao$^{1\dagger}$,
Michio Tateno$^{2}$,
Omar A. Saleh$^{1,2,3}$,\and
Zvonimir Dogic$^{1,3}$,
M. Cristina Marchetti$^{1,3\ast}$\and
	\small$^{1}$Department of Physics, University of California, Santa Barbara, CA 93106, USA\and
\small$^{2}$Department of Materials, University of California, Santa Barbara, CA 93106, USA\and
\small$^{3}$Interdisciplinary Program in Quantitative Biosciences, University of California, Santa Barbara, CA 93106, USA\and
\small$^{4}$Department of Physics, Initiative for Theory and Modeling of Living Systems, Emory University, Atlanta, Georgia 30322, USA\and
\small$^\ast$Corresponding author. Email: cmarchetti@ucsb.edu\and
\small$^\dagger$These authors contributed equally to this work.
}

\begin{document} 

\twocolumn[
\begin{@twocolumnfalse}
\maketitle
\begin{abstract}
We study phase separation between coexisting active and passive fluids in three-dimensions, using numerical simulation and experiments. Chaotic flows of the active phase drive giant interfacial deformations, causing the co-existing phases to interpenetrate and generate a continuously reconfiguring bicontinuous morphology which persists over the  lifetime of the active fluid.  Active bicontinuous structures are dominated by sheet-like interfaces, in marked difference from passive liquid-liquid phase separation which is controlled by saddle-like surfaces. Activity and surface tension control the  length scale of the bicontinuous structure. These results demonstrate how active stresses suppress the coarsening of conventional phase separation, generating steady-state reconfigurable morphologies not accessible with conventional surface-modifying agents or through quenching of transient phase separated structures.
\end{abstract}
\end{@twocolumnfalse}
]

\noindent
\section{Introduction}
Bicontinuous structures consist of two distinct interconnected and interpenetrating phases, where each phase spans the entire sample. By maximizing the interfacial area while retaining connectivity, bicontinuity underpins vital mechanical and chemical functionalities in diverse natural and synthetic systems.  For example, in living matter, the interconnected endoplasmic reticulum that spans the cell interior performs protein synthesis, lipid production, and molecular transport~\cite{holcman2018single,dayel1999diffusion}. The bone trabecular lattice balances the mechanical strength of the mineralized struts with nutrient transport through vascularized marrow channels~\cite{ liu2006quantification,herisson2018direct}. In synthetic systems, spinodally decomposed alloys derive exceptional strength from their interpenetrating nanoscale phases, while block copolymer gyroids form porous electrodes that optimize transport in batteries \cite{jinnai1995direct,zhang2011three,shi2021scaling,werner2018block}. 

A homogeneous fluid mixture spontaneously demixes into two distinct phases~\cite{cahn1958free}, a dynamical process that can generate a bicontinuous structure~\cite{loren2002determination,jinnai2003connectivity}. In systems in thermal equilibrium, however, bicontinuous structure is inherently transient as the two phases coarsen over time reaching bulk phase separation~\cite{siggia1979late,Bray2002}. %~\cite{siggia1979late,setru2021hydrodynamic,jinnai2003connectivity,scholten2005coarsening,groot1999role,bazazi2022spongy,araki2001three,tateno2021power}. 
Thus, preserving bicontinuity requires kinetic arrest of the coarsening. Established strategies include suppressing the fluidity of one of the two phases. This can be achieved by replacing one of the two fluids with an elastomer network \cite{fernandez2024elastic}, by polymerizing one of the phase-separating fluids during coarsening \cite{chan1996polymerization,lee2022elastomeric,muller2025cell}, or by vitrifying it through rapid cooling \cite{royall2018vitrification}. Alternatively, the interface itself can be stabilized by adsorbing a high density of colloids on the interface to form mechanically stable ``bijels''~\cite{stratford2005colloidal,herzig2007bicontinuous,huang2017bicontinuous}. Finally, block copolymers offer an alternate method for generating bicontinous structures through chemistry, but the scale of the resulting structure is controlled by the size of the constituent molecules~\cite{thomas1986ordered,hajduk1994gyroid,sai2013hierarchical}.

\begin{figure*}[h!]
\begin{center}
\includegraphics[width=0.9\textwidth]{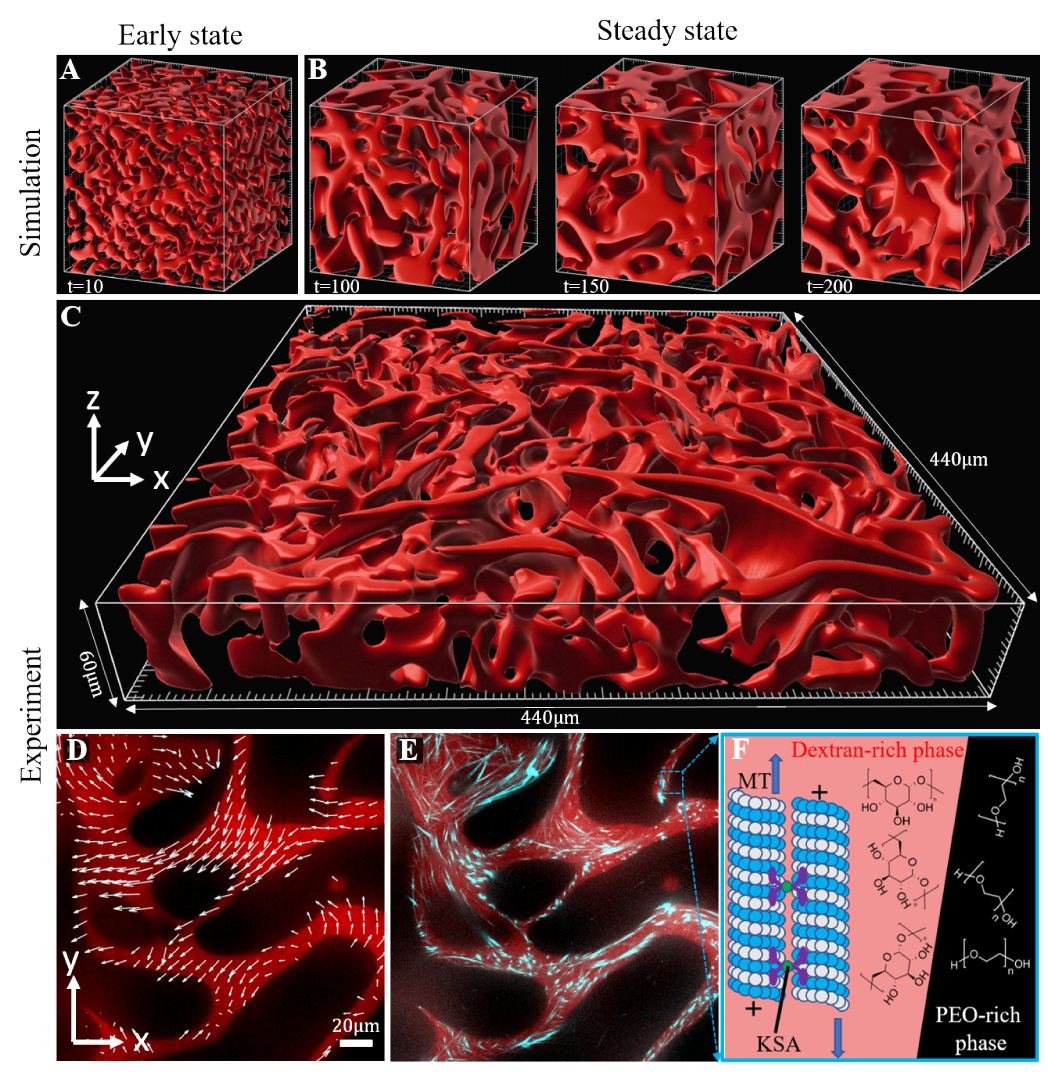}
\end{center}
\caption{\textbf{Active 3D phase separation yields a percolating network-like structure}. (A-B) Morphology of active phase separation in computer simulations from the early state to the steady state. Red represents the active phase ($\phi>0.5$). Simulation parameters: activity $|\alpha|=1.30,$ surface tension $\gamma=1.0,$ and active fraction $\phi_a = 0.40$. (C) Experimental morphology of the active phase (red) at steady state. (D) An $x$-$y$ cross-section of a phase-separating experimental sample. The white arrows indicate the active flows in the dextran phase (red), obtained from Particle Imaging Velocimetry. (E) MTs (cyan) partition into the dextran phase (red). (F) MTs in the dextran phase form bundles due to  the depletion effect. Kinesin clusters (KSA) bind to MT bundles driving their extension. The experimental samples (C-E) contain 1.8\% PEO and 1.8\% dextran, $\phi_a=0.37$, and are imaged at time $t=1.5$~h. The KSA concentration is 183~nM (C) and 92~nM (D, E). }
\label{setup}
\end{figure*}

Here we describe a fundamentally different pathway for generating robust and tunable bicontinuous structures that uses active fluids and relies on mechanical, rather than chemical, control of soft interfaces. Active fluids are composed of energy-consuming molecular units that collectively generate autonomous chaotic flows~\cite{aditi2002hydrodynamic,saintillan2008instabilities,sanchez2012spontaneous,wensink2012meso,zhou2014living}. Such flows and associated active stresses can power non-equilibrium dynamics and active assembly of passive colloids or fibers \cite{grober2023unconventional,berezney2024controlling}. They can also drive the deformation and actuation of soft interfaces and membranes~\cite{wu2000particle,keber2014topology,martinet2025emergent,takatori2020active,adkins2022dynamics,zhao2024asymmetric,sessa2026interfacial,sciortino2025active,laprade2025coarsening,liu2025nonequilibrium}. Here we show how active flows can transform conventional coarsening dynamics into a continuously reconfiguring bicontinuous active network, which persists for the lifetime of the active fluid and is stable over a wide range of volume fractions. The bicontinuous structure is controlled by activity and surface tension. The interfacial curvature of the active bicontinuous network is dominated by sheet-like geometries, in contrast to passive phase-separating fluids where bicontinuous structures are dominated by minimal saddle-like surfaces~\cite{cahn1958free,araki2001three, beysens1997kinetics, kwon2007coarsening, perrot1999morphology}. 

The three-dimensional geometry studied here is qualitatively different from previous work which explored the role of microtubule-based activity on phase separation of PEO/dextran mixtures. One class of studies explored active phase separation in very thin chambers where the sample was effectively two dimensional~\cite{adkins2022dynamics,sessa2026interfacial}. Another study examined the effect of activity on the shape and stability of a bulk-separated interface in a three dimensional sample~\cite{zhao2024asymmetric}. Here, in contrast, we explore the morphology of the late stage phase separated state and demonstrate the stability of bicontinuous structures.

% \begin{figure*}[h!]
% \begin{center}
% \includegraphics[width=0.9\textwidth]{Figures/setup.png}
% \end{center}
% \caption{\textbf{Active 3D phase separation yields a percolating network-like structure}. (A-B) Evolution of the morphology of active phase separation in 3D computer simulations from the early state to the steady state. Red represents the active phase ($\phi>0.5$). Simulation parameters: activity $|\alpha|=1.30,$ surface tension $\gamma=1.0,$ and active fraction $\phi_a = 0.40$. (C) Experimental morphology of the active phase (red) at steady state. The sample contains 92~nM KSA, 1.8\% PEO, 1.8\% dextran, and $\phi_a=0.38$, imaged at time $t=1.5$~h. (D) An $x$-$y$ cross-section of a phase-separating sample. The white arrows indicate the active flows in the dextran phase (red), obtained from Particle Imaging Velocimetry. (E) MTs (cyan) partition into the dextran phase (red). (F) MTs in the dextran phase form bundles due to  the depletion effect. Kinesin clusters (KSA) bind to MT bundles driving their extension.}
% \label{setup}
% \end{figure*}
 
\section{Results}

\subsection{Theoretical model of active phase separation}
We model an immiscible mixture of active and passive fluids in three dimensions within a continuum framework that couples Cahn-Hilliard phase separation to active nematodynamics. The concentration field $\phi(\mathbf{r})$ distinguishes the passive ($\phi=0$) and the active ($\phi=1$) phase, while the nematic tensor $\mathbf{Q}(\mathbf{r})$ captures the local orientational order within the active fluid. These fields are coupled through the incompressible Stokes flow $\mathbf{v}(\mathbf{r})$ and its spatial gradients (Methods~\ref{sec:continuummodel_sim})~\cite{caballero2022activity,zhao2024asymmetric,gulati2025active}.  

The phase separation between active and passive fluid is governed by the equilibrium surface tension $\gamma$. The sample composition is controlled by the active fraction $\phi_a = (1/V)\int d\mathbf{r}\; \phi(\mathbf{r})$, where $V = \int d\mathbf{r}$ is the total volume. The active stress tensor that drives Stokes flow is
\begin{equation}
\boldsymbol{\sigma}^{\text{active}} = \alpha \phi \mathbf{Q},
\end{equation}
where $\alpha <0$ is the extensile activity responsible for the turbulent-like dynamics of the active ($\phi=1$) phase. We choose a parameter regime in which the nematic tensor $\mathbf{Q}(\mathbf{r})$ relaxes to the isotropic state ($\mathbf{Q}=0$) in the absence of activity. Above a critical activity, active stresses destabilize the quiescent isotropic state and generate autonomous flows~\cite{putzig2016instabilities, srivastava2016negative,santhosh2020activity, vafa2021fluctuations}. 

In passive systems, the well-mixed state is unstable over a range of parameters, leading to a spinodal instability that generates spatially modulated structures that coarsen over time. Such dynamics reduces the interfacial energy, ultimately resulting in bulk phase separation~\cite{cahn1958free,siggia1979late}. Activity profoundly alters this process~\cite{cates2025active}. At high activity, turbulent-like flows stir the fluid and impede the thermodynamically driven phase separation~\cite{caballero2022activity, tayar2023controlling}. Intermediate activities do not suppress phase separation, but arrest coarsening and influence the morphology of the phase separated state. In two dimensions, moderate activity can generate percolating dynamical filamentary networks with a characteristic length scale controlled by the balance of activity and surface tension~\cite{gulati2025active}.  

We study the active-passive mixture at intermediate activities, using the numerical model. Starting with an initially uniform state, we observe the emergence of spatially modulated structures, which reach a length scale that plateaus over time (Fig.~\ref{setup}A, B, Movie~S1). The steady-state structures have a labyrinth like appearance, consisting of junction points that are connected by thread-like liquid bridges. The network is not static. The threads connecting different junctions appear to continuously rupture while new ones are generated. Such microscopic dynamics generates a ``living'' continuously reconfiguring network-like structure.

\subsection{Experimental realization of active phase separation} 
To experimentally study active-passive phase separation, we combined a previously described passive phase-separating polymer mixture of polyethylene oxide (PEO) and dextran with an active fluid (Methods~\ref{sec:experimental_design})~\cite{liu2012concentration,adkins2022dynamics,zhao2024asymmetric}. The activity is generated by microtubules (MTs) and streptavidin-bound clusters of kinesin molecular motors (KSA) (Fig.~\ref{setup}F). Both protein-based components partition into the dextran-rich phase (Fig.~\ref{setup}E). The kinesin clusters bind to adjacent MTs, transforming the chemical energy from adenosine triphosphate (ATP) hydrolysis into interfilament sliding and bundle extension, which in turn drives large-scale turbulent flows (Fig.~\ref{setup}D). 

Motivated by the theoretical prediction of bicontinuity, we studied active phase separation in three dimensions, using 100 $\mu$m thick chambers. The evolving structure was imaged with confocal microscopy. The active (dextran) and passive (PEO) regions are differentiated using fluorescence intensity (Methods~\ref{sec:data_analysis}). Internal activity generates large interfacial deformations that transform the morphology of the two coexisting phases, similar to what is observed in simulations.  Active flows generate a system spanning network-like structure (Fig.~\ref{setup} C). The network is continuously reconfigured by the active flows (Movie.~S2). Such dynamics persists for several hours, which is the active fluid lifetime (Fig.~\ref{fig:sup_evolution}). In contrast, a passive system lacking active components forms minority-phase droplets that coarsen over time (Fig.~\ref{fig:sup_passive}).

\begin{figure*}[h!]
    \centering
    \includegraphics[width=0.99\linewidth]{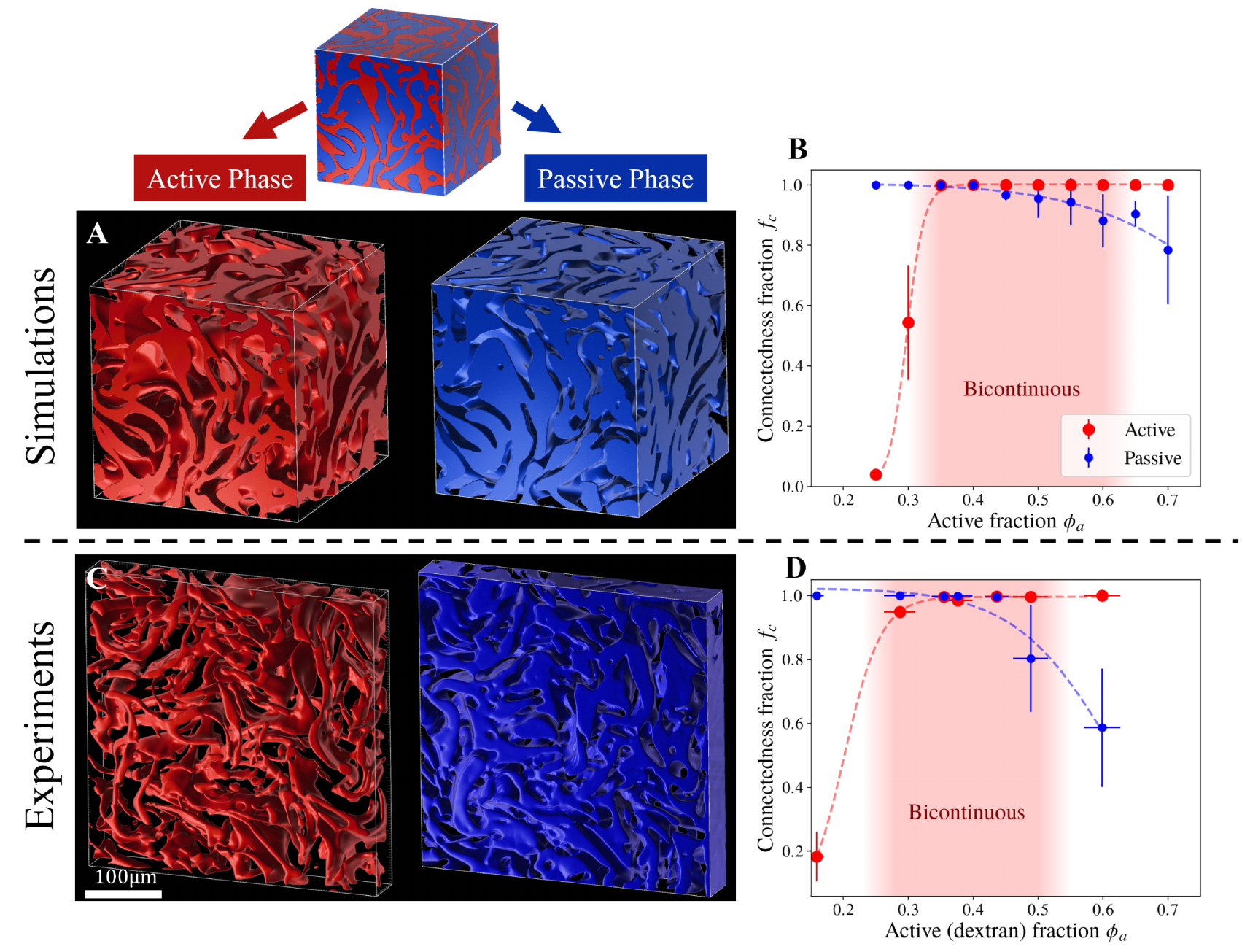}
    \caption{\textbf{Bicontinuous morphology persists over a wide range of volume fractions.} (A) Connected active (red) and passive phase (blue) in simulations, with $\phi_a = 0.40$ (B) Connectedness factor $f_c$ of the active and passive phases as a function of the volume fraction of the active phase $\phi_a$. The range where $f_c \approx 1$ for both phases highlighted in pink indicates bicontinuity. Simulation parameters: $|\alpha| = 0.70, \gamma=1.0$. (C) Morphology of the connected active (red) and passive phase (blue) in the experiment. $\phi_a = 0.37$. (D) The connectedness fraction $f_c \approx 1$ for intermediate values of $\phi_a$, indicating bicontinuity. Experimental samples initially contain 183~nM KSA, 1.8\% PEO, and 1.8\% dextran.}
    \label{fig:bicontinuous}
\end{figure*}

\begin{figure*}[h!]
    \centering
    \includegraphics[width=0.99\linewidth]{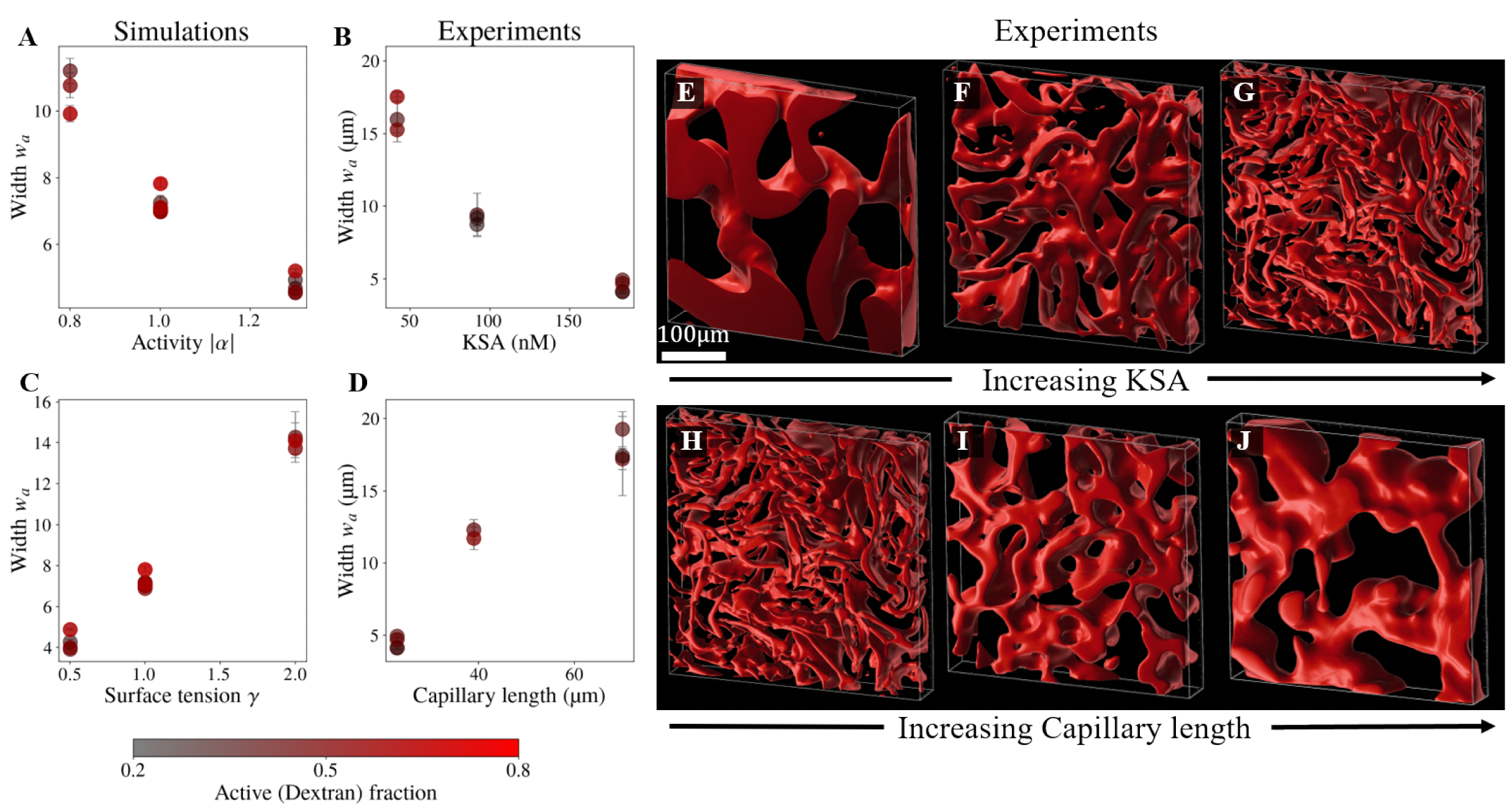}
    \caption{\textbf{Active width $w_a$ decreases with activity and increases with surface tension}. (A-B) $w_a$ decreases with activity $|\alpha|$ in simulations and with KSA concentration in experiments. (C-D) $w_a$ increases with surface tension $\gamma$ in simulations and with capillary length in experiments. In (A-D) the color of the dots represents the volume fraction $\phi_a$ of the active phase (see color bar) and shows that $w_a$ is essentially independent of $\phi_a$. (E-G) 3D visualization of the dextran phase (red) in experiments with increasing KSA concentration, showing a corresponding decrease of $w_a$.  (H-J) 3D visualization of the dextran phase (red) in experiments with increasing capillary length, showing a corresponding increase of $w_a$. The experimental samples (B, E-G) contain 1.8\% PEO and 1.8\% dextran. KSA concentrations are increased from 42~nM (E), 92~nM (F), to 183~nM (G). Samples (D, H-J) contain 183~nM KSA. PEO and dextran concentrations are both increased from 1.8\% (H), 1.9\% (I), to 2.0\% (J). 
    % \PGcom{Distinguish steady state vs fixed long time in simulations vs experiments}
    }
    \label{fig:width}
\end{figure*}

\subsection{Characterizing the bicontinuous morphology}
Numerical simulations and experiments demonstrate that 3D active phase separation generates a steady-state dynamical structure. Active and passive phases form distinct yet intercalated system-spanning networks. Taken together, they generate a bicontinuous structure (Fig.~\ref{fig:bicontinuous} A,C). To analyze the bicontinuity in simulation, we threshold the 3D structure into passive and active regions, corresponding to $\phi<0.5$ and $\phi>0.5$, respectively. The connectedness fraction, $f_c$, of each phase is defined as the volume of the largest connected active/passive component divided by the total volume of that component~\cite{gulati2025active}. 

To determine the range of bicontinuity, we vary the fraction of the active phase $\phi_a$, while keeping the activity and surface tension fixed. Bicontinuous structures, defined as those where $f_c\approx 1$ for both phases, are observed from $0.3 < \phi_a < 0.6$ (Fig.~\ref{fig:bicontinuous} B) in simulation. For $\phi_a< 0.3$, bicontinuity is broken as the active domains disintegrate into isolated droplets. For $\phi_a> 0.6$, the passive domains are broken up by active flows and become disconnected. The range of stable bicontinuous structures is controlled by activity and surface tension (Fig.~\ref{fig:sup_bicontinuity_activities}). For example, doubling the activity generates stable bicontinuous structures over the entire regime of phase separation $0.0 < \phi_a < 1.0$.  

Motivated by these predictions, we have explored the phase space of stable bicontinuous networks in experiments. We use a phase recombination method to keep the polymer concentration (surface tension) and the motor protein concentration (activity) fixed within each phase, while the active fraction $\phi_a$ varies between 0.16-0.60. We (1) prepare an initial mixture with controlled concentrations of PEO, dextran and KSA, then (2) centrifuge the sample to generate bulk phase separation, and (3) recombine the two separated phases into a new mixture with chosen $\phi_a$ (Methods~\ref{sec:experimental_design}).  

In experiments, the bicontinous structure for the particular is observed over the range of volume fractions $0.3<\phi_a<0.5$ (Fig.~\ref{fig:bicontinuous}D). For active fractions greater than $0.5$, the passive phase disintegrates into droplets, while for fractions less than $0.3$ the active phase breaks up into droplets, significantly decreasing $f_c$. Notably, the range of stable bicontinuous structures is not centered around $\phi_a=0.5$: bicontinuity persists for a low volume fraction of the active phase $\phi_a \approx 0.3$. In comparison, it disappears as soon as the active phase reaches the majority $\phi_a \approx 0.5$. This asymmetry suggests that the active phase containing MTs remains connected at a very low volume fraction, while the passive PEO-rich/MT-devoid phase more easily breaks apart into the droplets. This asymmetry is more pronounced in experiments than simulations. Intriguingly, similar asymmetry was observed in the directional invagination and break-up of a bulk three-dimensional active interface, where it was attributed to the nematic elasticity of the active fluid~\cite{zhao2024asymmetric}.  

\subsection{Controlling the morphology: active width}
Simulations show that the morphology of the bicontinuous network is controlled by activity and surface tension. To characterize this structure, we measure the width $w_a$ of the active network, defined as the volume to surface ratio of the largest connected active domain. We find that $w_a$ increases with decreasing activity and with increasing equilibrium surface tension. Furthermore, $w_a$ is essentially independent of the volume fraction $\phi_a$  (Fig.~\ref{fig:width}A, C), suggesting that intrinsic active processes control the length scale of the bicontinuous morphology.

To experimentally test these predictions, we vary the KSA concentration, which is a proxy for activity, and the PEO/dextran concentration, which controls the surface tension. Due to a density difference, the dextran phase gradually sediments due to gravity over tens of minutes, causing a variation in $w_a$ (Fig.~\ref{fig:gravity}). To minimize this effect, we continuously invert the top and bottom of the chamber by slowly rotating the samples. The sample is then moved from the rotation stage to the microscope and imaged for a maximum duration of 5 minutes (Methods~\ref{sec:experimental_design}). The effective surface tension is estimated from measurements of the capillary length for different PEO/dextran concentrations (Fig.~\ref{fig:sup_capillary}). The experimental trends follow the predictions of the numerical model: increasing surface tension increases $w_a$; increasing activity decreases $w_a$ (Fig.~\ref{fig:width} A-D). Furthermore, $w_a$ is largely independent of the volume fraction. 

The behavior of the width $w_a$ suggests that the morphology is controlled by two competing mechanisms: the stretching of the interface due to extensile active flows and its relaxation controlled by surface tension. Extensile active forces are known to  align the nematic director with the interface~\cite{blow2014biphasic, coelho2023active, zhao2024asymmetric}. This alignment then drives active shear flows that tangentially stretch the interface. The stretching is countered by surface tension that favors minimizing the interface. In general we expect activity to also modify the surface tension, as shown for instance in scalar models of active phase separation~\cite{cates2025active}. Here we assume that the main effect of activity is to build up local nematic order and align the liquid crystal with the interface, and use the bare (passive) surface tension in the scaling arguments below. In the isotropic liquid crystalline phase, the active stretching rate can be approximated as 
\begin{equation}
\tau^{-1}_{\text{active}} \sim \frac{S(|\alpha| -\alpha_c)}{\eta} \sim \frac{(|\alpha| -\alpha_c)^{3/2}}{\eta}\;,
\end{equation}
where $\eta$ is the fluid viscosity, $S$ is the scalar nematic order parameter and $\alpha_c$ is the threshold above which activity drives local nematic order \cite{zhao2024asymmetric,gulati2025active} (Methods~\ref{sec:continuummodel_sim}). We approximate the scalar order parameter with the mean value generated by activity, $S \sim \sqrt{(|\alpha|-\alpha_c)}$ \cite{santhosh2020activity, giomi2012banding}. In three-dimensions, the rate at which the interface relaxes due to surface tension is given by
\begin{equation}
\tau^{-1}_{\text{diffusive}} \sim \frac{\gamma}{V}\;,    
\end{equation}
where $V$ is the total volume of the active domain. We write this as $V = w_a^{{d_f}}\ell^{\left(3-{d_f}\right)},$ where $w_a$ is the width of the active region , $\ell$ is set by mass conservation, and $d_f < 3$ is controlled by the active process ~\cite{gulati2025active}. Tube-like filamentary structures correspond to $d_f=2$ whereas sheet-like structures, with one short axis, correspond to $d_f=1$. At steady state, the active width is determined by balancing active stretching and interfacial relaxation,
$\tau_{\text{active}}^{-1} \sim  \tau_{\text{diffusive}}^{-1}$, with the result
\begin{equation}
    w_a \sim \left(\dfrac{\gamma}{(|\alpha|-\alpha_c)^{3/2}}\right)^{1/d_f} = \left(\dfrac{|\alpha|-\alpha_c}{\gamma^{2/3}}\right)^{-3/2d_f}\;. \label{eq:scalingLaw}
\end{equation}

\begin{figure}[h!]
    \centering
    \includegraphics[width=0.95\linewidth]{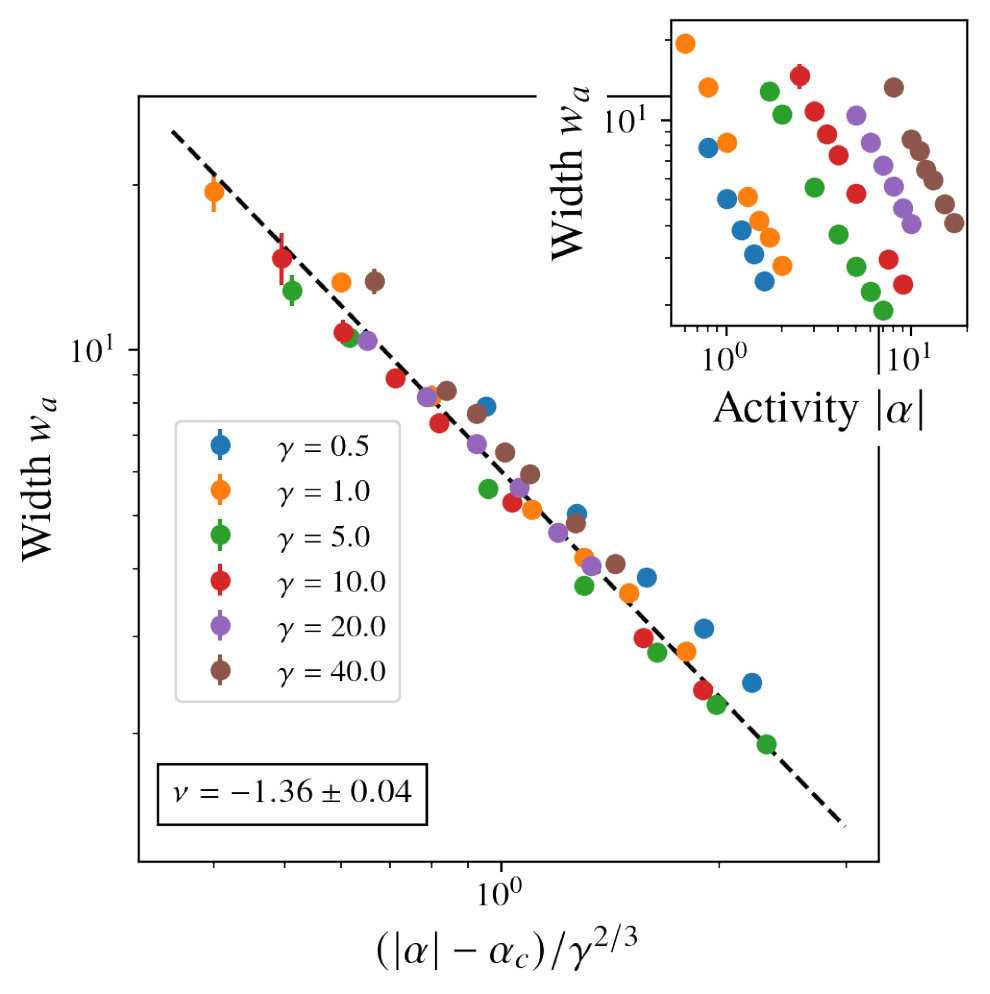}
    \caption{\textbf{Scaling law for the width $w_a$ in simulations.} Steady state width $w_a= \text{Volume}/\text{Surface Area}$ of the connected active domain in the bicontinuous region as a function of activity $|\alpha|$ and equilibrium surface tension $\gamma$, for a fixed active fraction $\phi_a=0.40$.  The main figure shows that the data  collapse when plotted versus $(|\alpha|-\alpha_c)/\gamma^{2/3}$  (Eq.~\ref{eq:scalingLaw}). The black dashed line shows the best fit (via linear regression) to $w_a \sim (|\alpha|-\alpha_c)/\gamma^{2/3})^\nu$ for $\nu = -1.36 \pm 0.04$. The inset shows the unscaled active width as a function of activity, for various values of  $\gamma$. Fixed parameters: $K = 24.0, \alpha_c = 2\eta r/\lambda \Gamma  =0.2$. }
    \label{fig:widthScaling}
\end{figure}

Using simulations, we have measured the active width as a function of both activity and surface tension. The collapse of the data  supports a scaling law $w_a  \sim \left( (|\alpha| - \alpha_c)/\gamma^{2/3}\right)^{\nu},$ with $\nu \approx -1.36$, corresponding to $d_f = -3/(2\nu) \approx 1.1$~(Fig.~\ref{fig:widthScaling}). This scaling suggests that the bicontinuous morphology is dominated by sheet-like active domains, whose width is governed by the balance of active stretching and passive relaxation.

\subsection{Characterizing curvature and morphology}
\begin{figure*}[h!]
\begin{center}
\includegraphics[width=0.99\textwidth]{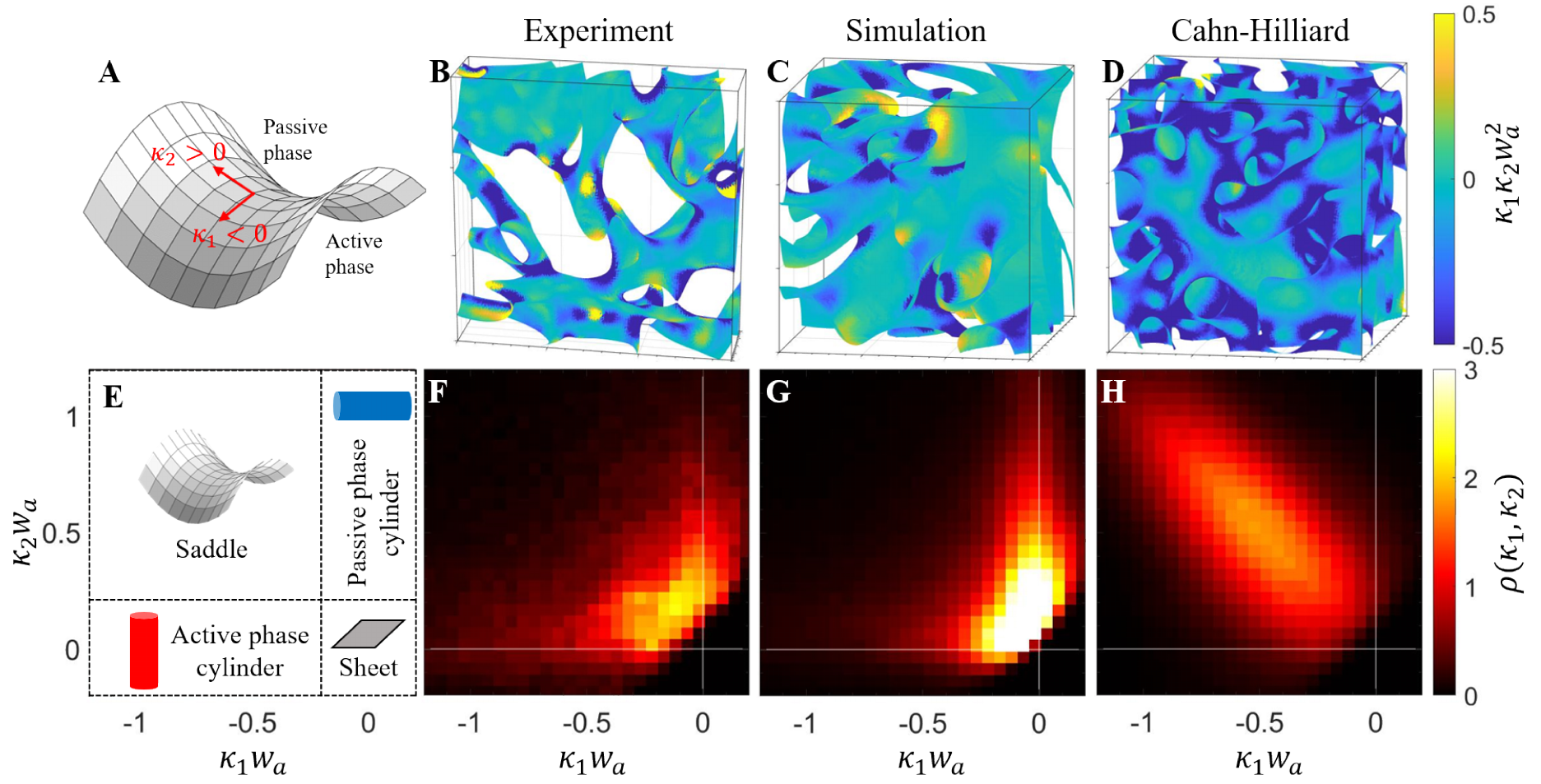}
\end{center}
\caption{\textbf{Comparing the interface morphology in passive and active bicontinuous structures}. (A) The two principal curvatures $\kappa_1$ and $\kappa_2$ are calculated everywhere on the interface. (B-D) 3D visualization of the normalized Gaussian curvature $\kappa_1\kappa_2w_a^2$ of the interface from experiment (B), simulations of the active/passive mixture(C) and simulations of the passive Cahn-Hilliard equation (D). The color indicates the value of the Gaussian curvature. (E-H) Distribution $\rho(\kappa_1,\kappa_2)$ of normalized principal curvatures $\kappa_1w_a$ and $\kappa_2w_a$. Different values of $\kappa_1$ and $\kappa_2$ correspond to different geometries (E). $\rho$ peaks at small values of $\kappa_1$ and $\kappa_2$ in experiments (F) and simulations (G), indicating sheet-like interfaces. $\rho$ peaks at $\kappa_1 \ll 0, \kappa_2 \gg 0$ in simulations of the Cahn-Hilliard equation (H), indicating a saddle-like interface. The experimental sample (B,F) contains 92~nM KSA, 1.8\% PEO, 1.8\% dextran, and $\phi_a=0.37$. Simulation parameters (C,G): $|\alpha|=1.30, \gamma=1.0,$ and  $\phi_a=0.40$. }
\label{fig:curvature}
\end{figure*}

The scaling argument described above suggests the dominance of sheet-like structures. To investigate further, we quantify the surface curvature in both experiment and simulation. We compute the values of the two principal curvatures $\kappa_1(\mathbf{s})$ and $\kappa_2(\mathbf{s})$, where $\mathbf{s}$ is a point on the interface $\mathcal{S}$. We define $\kappa_1(s)<\kappa_2(s)$ for each $s$ (Fig.~\ref{fig:curvature}A) (Methods~\ref{sec:data_analysis}). We visualize the interface structure with its normalized Gaussian curvature $w_a^2 \kappa_1(\mathbf{s}) \kappa_2(\mathbf{s})$ for $\mathbf{s} \in \mathcal{S},$ where $w_a$ is the width of the active phase, as defined in the previous section. In both experiments and simulations, the interface generated by active phase separation is dominated by regions of normalized Gaussian curvature close to zero (Fig.~\ref{fig:curvature} B,C). We contrast this with the transient morphology obtained in passive phase separation from the Cahn-Hilliard equation, which is dominated by saddle-like surfaces with negative Gaussian curvature~\cite{cahn1958free,araki2001three, kwon2007coarsening} (Fig.~\ref{fig:curvature}D).

From the principal curvature values, we compute the joint probability distributions $\rho(\kappa_1, \kappa_2)$. Different regions in the $\kappa_1$-$\kappa_2$ plane indicate different structures (Fig.~\ref{fig:curvature} E). In both experiments and simulations, the joint distribution is peaked at small values of both $\kappa_1$ and $\kappa_2,$ which corresponds to sheet-like domains (Fig.~\ref{fig:curvature} F,G). The vertical and horizontal branches centered around zero-curvature corresponds to cylinder-like surfaces, which are associated with the sheet edges. In comparison, the probability distribution obtained from the Cahn-Hilliard equation is characterized by large positive-negative pairs of principal curvatures correspond to saddle-like surfaces (Fig.~\ref{fig:curvature}H). The curvature analysis supports the finding that the connected active domain is dominated by sheet-like structures.

% \begin{figure*}[h!]
% \begin{center}
% \includegraphics[width=0.99\textwidth]{Figures/fig5.png}
% \end{center}
% \caption{\textbf{Comparing the interface morphology in passive and active bicontinuous structures}. (A) The two principal curvatures $\kappa_1$ and $\kappa_2$ are calculated everywhere on the interface. (B-D) 3D visualization of the normalized Gaussian curvature $\kappa_1\kappa_2w_a^2$ of the interface from experiment (B), simulations of the active/passive mixture(C) and simulations of the passive Cahn-Hilliard equation (D). The color indicates the value of the Gaussian curvature. (E-H) Distribution $\rho(\kappa_1,\kappa_2)$ of normalized principal curvatures $\kappa_1w_a$ and $\kappa_2w_a$. Different values of $\kappa_1$ and $\kappa_2$ correspond to different geometries (E). $\rho$ peaks at small values of $\kappa_1$ and $\kappa_2$ in experiments (F) and simulations (G), indicating sheet-like interfaces. $\rho$ peaks at $\kappa_1 \ll 0, \kappa_2 \gg 0$ in simulations of the Cahn-Hilliard equation (H), indicating a saddle-like interface. The experimental sample (B,F) contains 92~nM KSA, 1.8\% PEO, 1.8\% dextran, and $\phi_a=0.38$. Simulation parameters (C,G): $|\alpha|=1.30, \gamma=1.0,$ and  $\phi_a=0.40$. }
% \label{fig:curvature}
% \end{figure*}

\section{Discussion}
We showed how activity transforms the coarsening domains of passive three-dimensional phase separation into a continuously reconfiguring bicontinuous structure that maintains its morphology. The bicontinuous morphology is observed across a wide range of volume fractions, while its width is controlled by activity and surface tension. The active bicontinuous structure is composed of sheet-like domains, in contrast to saddle-like interfaces observed in passive phase separation~\cite{cahn1958free,araki2001three, beysens1997kinetics, kwon2007coarsening, perrot1999morphology}.

Extensive previous work examined the effect on activity on phase separation has been explored from multiple perspectives. In one regime, also studied here, active isotropic or nematic liquid crystals coexist with a passive fluid~\cite{kempf2019active,adkins2022dynamics, zhao2024asymmetric,gulati2024traveling,sessa2026interfacial, laprade2025coarsening}. In such systems, active stresses are present throughout the bulk of the active fluid. System-spanning active networks over a wide range of volume fractions have been studied in two-dimensional simulations~\cite{gulati2025active}. Building on these findings, we use simulations and experiments to demonstrate that bicontinuity is a ubiquitous feature of three-dimensional active phase separation. An important ingredient required for the formation of bicontinuous structures is active anchoring~\cite{blow2014biphasic}, where the active units align along the interfaces to generate autonomous flows that stretch the active phase into sheet-like domains. This elongation continuously connects separate droplets into a network and dynamically counters the collapsing effect of surface tension. Theoretical work also explored a different regime, described by scalar models of active/passive mixtures, where the mixture is described in terms of the conserved concentration field, coupled to flow~\cite{tiribocchi2015active,singh2019hydrodynamically,cates2025active}. In this regime, active stresses are driven by gradients of the concentration field that are localized at the interface. Whether bicontinuous structures can be obtained in three-dimensional scalar models, perhaps as a result of active renormalization of surface tension, remains, to our knowledge, an open question.  

The structures in active phase separation are determined by the competition between active stress and surface tension. We have shown that this competition generates bicontinuous morphology and controls the width of the active sheets. The bicontinuity, however, is only observed over a wide but well-defined range of activities. On the one hand, when activity is very small compared to surface tension, active flows cannot elongate the droplets, but only induce small interfacial deformations. In this case, one phase irreversibly turns into droplets through a Plateau-Rayleigh instability~\cite{siggia1979late}. On the other hand, for very high activities, the energy input overcomes the effect of the thermodynamic chemical potential, suppressing the phase separation itself and generating a uniform mixture \cite{caballero2022activity,tayar2023controlling}. Furthermore, bicontinuous structures are only observed in a finite range of active fraction $\phi_a$ (Fig.~\ref{fig:bicontinuous}). At low fractions of the active fluid elongated droplets cannot merge. In this regime, the system transitions from droplets to a uniform phase with decreasing surface tension, without an intermediate bicontinuous state~\cite{tayar2023controlling}. Finally, gravity causes a non-uniform distribution of the two phases, generating extreme densities $\phi_a$ which do not allow bicontinuity~\cite{zhao2024asymmetric}. Our work provides specific techniques and parameters to overcome these difficulties and achieve a controllable bicontinuous fluid.

Bicontinuous structures also arise in passive phase separation~\cite{tateno2021power}. However, without active forces, passive restorative interfacial tension drives the system to a state with a minimal surface area. The characteristic length of conventional coarsening dynamics grows with a power law $w\propto t^{\nu_p}$, where $\nu_p$ varies between 1/3 and 1~\cite{cahn1958free,siggia1979late,tateno2021power}. Therefore, bicontinuous structures in passive systems are transient. In contrast, the active stresses maintain a dynamical, bicontinuous steady state throughout the entire active fluid lifetime.

Biological systems are adaptable and self-healing. Using continuous energy input, they maintain dynamical steady states, adapt to damage, and resist degenerative processes. Our work demonstrates how using continuous energy input can generate controllable and highly dynamical biomimetic structures and morphologies that are not easily achievable in equilibrium. 

\section{Materials and Methods}
\subsection{Continuum model and simulation details}
\label{sec:continuummodel_sim}
We use a continuum model of phase separation in a mixture of an active liquid crystal and  passive fluid. The model couples Cahn-Hilliard dynamics to the hydrodynamic of a nematic liquid crystal. It is formulated in terms of coupled equations for the phase fraction $\phi$ of active fluid, the nematic order parameter $\mathbf{Q}$ and fluid flow $\mathbf{v}$, given by  
\begin{equation}
\begin{aligned}
    \partial_t \phi + \mathbf{v}\cdot \boldsymbol{\nabla}\phi = M\nabla^2\mu,\\
    \partial_t \mathbf{Q} + \mathbf{v}\cdot \boldsymbol{\nabla} \mathbf{Q} + [\boldsymbol{\Omega}, \mathbf{Q}] = \lambda \mathbf{A} + \frac{1}{\Gamma}\mathbf{H},\\
    0 = \eta \nabla^2 \mathbf{v} - \boldsymbol{\nabla} P + \boldsymbol{\nabla} \cdot \boldsymbol{\sigma},
\end{aligned}\label{eq:continuumModel}
\end{equation}
where $\phi(\mathbf{r}) \in [0,1]$. The Stokes flow $\mathbf{v}(\mathbf{r})$ and its gradients given by the strain rate $A_{ij} =  (\partial_i v_j + \partial_j v_i)/2$ and the vorticity tensor $\Omega_{ij} = (\partial_i v_j -\partial_j v_i)/2$ couple $\phi$ and $\mathbf{Q}$.

Phase separation between the active $(\phi=1)$ and the passive fluid $(\phi=0)$ is driven by the chemical potential $\mu = \delta F_\phi/\delta \phi,$ where 
%$F_\phi$ corresponds to a symmetric double well potential given by 
$F_\phi = (3\gamma/\xi)\int d\mathbf{r}~ \phi^2 (\phi-1)^2 +\xi^2 (\nabla \phi)^2/2$. Here $\gamma$ is the equilibrium surface tension and $\xi$ is the width of the equilibrium interface between the two phases. We characterize the composition of the fluid mixture using the (conserved) active fraction $\phi_a = \int d\mathbf{r} \;\phi(\mathbf{r}).$

The (traceless) nematic tensor $\mathbf{Q}$ captures the local nematic orientational order inside the active fluid. The molecular field $H_{ij} = \left(\delta F_Q/\delta Q_{ij}\right)^{\rm{S,T}}$ controls the relaxation of the nematic tensor via the Landau-de Gennes free energy $F_Q = \int d\mathbf{r} \; r \Tr\{\mathbf{Q}^2\}/2 +w \Tr\{\mathbf{Q}^3\}/3 + u\Tr\{\mathbf{Q}^2\}^2/4 + K (\nabla \mathbf{Q})^2/2,$ with $r, u>0.$

The nematic field is defined everywhere, but in the absence of activity it relaxes to $\mathbf{Q} =0$ since $r>0$. Due to the flow alignment and vorticity couplings, self-generated flows result in non-zero local nematic order within the active ($\phi=1$) phase, as we discuss below.

The flow is governed by the Stokes equation with incompressibility, i.e, $\boldsymbol\nabla \cdot \mathbf{v} =0$ enforced via the pressure $P$. The stress tensor $(\boldsymbol{\sigma} = \boldsymbol{\sigma}^c + \boldsymbol{\sigma}^a)$ consists of the capillary stress ($\boldsymbol{\nabla}\cdot \boldsymbol{\sigma}^{c} = -\phi \boldsymbol{\nabla} \mu$), and the active stress, $\boldsymbol{\sigma^{a}}$, proportional to the phase field $\phi$ and the local nematic order $Q_{ij}$ i.e. 
\begin{equation}
    \boldsymbol\sigma^{a} = \alpha \phi \mathbf{Q}\;,
\end{equation}
with $\alpha<0$ corresponding to extensile activity. Within the active phase ($\phi=1$) self generated flows arise above a critical activity $\alpha_c = 2\eta r/\lambda \Gamma$. For $|\alpha| > \alpha_c,$ the positive feedback of flow alignment due to the active flows generated by perturbations of the nematic field leads to active turbulence \cite{santhosh2020activity, vafa2021fluctuations, putzig2016instabilities, srivastava2016negative}. 

Unless otherwise specified, the parameter values are reported in time units of $\Gamma/u$, length units of $\sqrt{ M \Gamma}$ and energy density in units of $u$. Throughout this paper, we fix the flow alignment parameter $\lambda =1.0$, viscosity $\eta=1.0$, interface width $\xi =\sqrt{3/2}$ and free energy parameters $r=0.1, w=0$.\footnote{We only consider large nematic order generated by active stresses, away from the isotropic to nematic transition, and hence can safely ignore the cubic term that appears in 3D. We verify in simulations that including a non-zero cubic energy contribution does not affect the stability of the bicontinuous phase.} We consider different values of the nematic elasticity $K,$ and systematically vary the activity $|\alpha|,$ the equilibrium surface tension $\gamma,$ and the active volume fraction $\phi_a.$ The simulations are done with periodic boundary conditions and performed using self-developed pseudo spectral solvers \cite{caballero2024cupss} on a uniform three-dimensional grid of size $170\times 170\times 170,$ typically for $10^5$-$10^6$ time steps until at least the system reaches steady state. 
%(see SI Fig~\ref{}). 

\subsection{Experimental Design}
\label{sec:experimental_design}
The experimental system of active-passive phase separation is based on previous protocols~\cite{adkins2022dynamics,zhao2024asymmetric}, with a decreased concentration of coexisting polymers to reduce surface tension. PEO (100 kDa, Sigma-Aldrich) and dextran (450 to 650 kDa, Sigma-Aldrich) were added into M2B buffer (80 mM K-pipes, 2 mM MgCl$_2$, 1 mM EGTA, pH 6.8). We studied three initial concentrations: 1.8\% PEO and 1.8\% dextran, 1.9\% PEO and 1.9\% dextran and 2.0\% PEO and 2.0\% dextran. Decreasing polymer concentration decreased the capillary length and surface tension (Fig.~\ref{fig:sup_capillary}). For visualization, we added ($<$0.1\% w/w) of amino-dextran (Fina Biosolutions) fluorescently labeled with Alexa-Fluor 488 NHS Ester (Invitrogen).

Tubulin was purified and labeled as described previously~\cite{tayar2022assembling}. A minor tubulin fraction (3\%)  was conjugated to AZDye 647 NHS Ester (Fluoroprobes) using a succinimidyl ester linkage. Labeled and unlabeled tubulin monomers were co-polymerized into microtubules (MTs) at 37 $^\circ$C using 0.6 mM guanylyl-(alpha,beta)-methylene-diphosphonate (GMPCPP, Gena Biosciences) and 1.2 mM Dithiothreitol (DTT, Gena Biosciences) in M2B buffer \cite{tayar2022assembling}. The MTs were sonicated for 60 seconds in a Branson 1800 sonicator to reduce filament length, which generated softer bundles that fragment more easily. The final MT concentration was 0.67~mg/mL. Kinesin-401 protein fused to a biotin-carboxyl carrier domain was expressed and purified using established protocols~\cite{tayar2022assembling}. Kinesin-401 was combined with tetrameric streptavidin in an M2B buffer with 5~mM DTT, forming a kinesin-streptavidin cluster (KSA). The final concentrations of KSA in the samples varied from 42-183~nM.

The PEO-dextran mixture was mixed together with MTs, KSA, caged ATP (NPE-caged adenosine triphosphate, P3-(1-(2-nitrophenyl)ethyl) ester, disodium salt, 1~mM final), antioxidants, and an ATP regeneration system made according to established protocols \cite{tayar2022assembling}. The antioxidants contained (in final concentrations) 2~mM Trolox, 5~mM DTT, 3.3~mg/mL glucose, 200~$\mu$g/mL glucose oxidase, and 35~$\mu$g/mL catalase dissolved in a phosphate buffer (20~mM K$_2$HPO$_4$ and 100~mM KCl in DI water, pH 7.4). The ATP regeneration system contained (in final concentrations) 26~mM phosphoenol Pyruvate (PEP) (Beantown Chemical) and pyruvate kinase/lactate dehydrogenase (Sigma-Aldrich) dissolved in M2B. 

During each experiment, a large 48~$\mu$L mixture containing the above ingredients was made in an Eppendorf tube. The sample was then centrifuged at 2000~$g$ relative centrifugal force for 1 min (Fisher Scientific, 05090-128) to generate bulk separated lighter PEO-rich and dense dextran-rich phases with a clear interface. MTs and KSA partitioned into the dextran phase. The two separated phases were separated into two tubes. Predetermined amounts from pure PEG and Dextran phases were mixed into a new 8~$\mu$L sample. The fraction of the dextran phase in this $8~\mu$L sample is the active fraction $\phi_a$. This remixing procedure ensured that the concentration of all components within each phase remained the same while $\phi_a$ varied between 0.15-0.60 (Fig.~\ref{fig:bicontinuous}, Fig.~\ref{fig:width}). The whole procedure was performed in a dark room to avoid releasing the caged ATP and activating the sample.

The final active sample was vortex-mixed for 10 s (Genie-2 Vortex Mixer) and placed within the chamber constructed by flanking the parafilm with two Polyethylene glycol (PEG) coated glass coverslips. The coverslips were coated with 10\% (w/w) mPEG-silane (20~kDa, Biopharma) dissolved in anhydrous Dimethyl sulfoxide (DMSO) \cite{gidi2018efficient}. The constructed chamber had a dimension of 4mm$\times$18mm$\times$0.1mm. The chamber was sealed with adhesive glue (Norland) and the glue was polymerized through 30-second exposure to ultraviolet light. The UV exposure released the caged ATP initiating activity. 

The samples were imaged with a spinning disk confocal microscope (X-Light V2, CrestOptics) and a sCMOS camera using 2x binning (Prime95B, Photometerics). Samples with 2.0\% PEO, 2.0\% dextran and samples with 42~nM KSA were imaged using a 10x air immersion objective (NA=0.5) with a 2~$\mu$m z-step. Other samples were imaged with a 40x water immersion objective (NA=1.15) with a 1~$\mu$m z-step. Samples with initially 1.8\% PEO, 1.8\% dextran and 183~nM KSA were imaged in dextran and MT channels and had a 14~s time interval. Other samples were imaged only in the dextran channel and had a 7~s time interval. 

The denser dextran phase sediments in the $-z$ direction over tens of minutes. This would generate two bulk phase separated phases, with the active phase perforated with passive droplets at the bottom of the chamber. To minimize this effect, samples were placed on a rotator (Benchmark Scientific H2020 Roto-Therm) and rotated at 24~rpm at room temperature for 1.5 hours. This reduced sedimentation (Fig.~\ref{fig:sup_evolution}). For imaging, the samples were transferred to the microscope and imaged for a maximum of 5 minutes (Fig.~\ref{fig:gravity}).

\subsection{Data Analysis}
\label{sec:data_analysis}
The experimental data were analyzed using established methods \cite{zhao2024asymmetric}. Confocal microscopy images were first classified into the dextran-rich phase and PEO-rich phase with the machine-learning software Ilastik based on intensity thresholding \cite{berg2019ilastik}. The interface was extracted between the two phases to calculate the width of the structure (Fig.~\ref{fig:width}). The interface was then transformed into a triangular mesh form using the software MeshLab \cite{cignoni2008meshlab}. The curvature was then calculated using published codes in MATLAB\cite{gut_matlab,rusinkiewicz2004estimating}. 3D visualization of both simulation and experiment data was done with software Imaris   (Fig.~\ref{setup},\ref{fig:bicontinuous},\ref{fig:width})\cite{Imaris}. 

%%%%%%%%%%%%%%%% REFERENCES %%%%%%%%%%%%%%%

\clearpage % Clear all remaining figures and tables then start a new page

\bibliography{ref} 

@article{holcman2018single,
  title={Single particle trajectories reveal active endoplasmic reticulum luminal flow},
  author={Holcman, David and Parutto, Pierre and Chambers, Joseph E and Fantham, Marcus and Young, Laurence J and Marciniak, Stefan J and Kaminski, Clemens F and Ron, David and Avezov, Edward},
  journal={Nature cell biology},
  volume={20},
  number={10},
  pages={1118--1125},
  year={2018},
  publisher={Nature Publishing Group}
}

@article{martinet2025emergent,
  title={Emergent dynamics of active elastic microbeams},
  author={Martinet, Q and Li, YI and Aubret, A and Hannezo, E and Palacci, J},
  journal={Physical Review X},
  volume={15},
  number={4},
  pages={041017},
  year={2025},
  publisher={APS}
}

@article{takatori2020active,
  title={Active contact forces drive nonequilibrium fluctuations in membrane vesicles},
  author={Takatori, Sho C and Sahu, Amaresh},
  journal={Physical Review Letters},
  volume={124},
  number={15},
  pages={158102},
  year={2020},
  publisher={APS}
}

@article{keber2014topology,
  title={Topology and dynamics of active nematic vesicles},
  author={Keber, Felix C and Loiseau, Etienne and Sanchez, Tim and DeCamp, Stephen J and Giomi, Luca and Bowick, Mark J and Marchetti, M Cristina and Dogic, Zvonimir and Bausch, Andreas R},
  journal={Science},
  volume={345},
  number={6201},
  pages={1135--1139},
  year={2014},
  publisher={American Association for the Advancement of Science}
}

@article{laprade2025coarsening,
  title={Coarsening of biomimetic condensates in a self-stirring active fluid},
  author={Laprade, Jeremy and Frechette, Layne and Amey, Christopher and Cusi, Adrielle and Baskaran, Aparna and Rogers, W Benjamin and Duclos, Guillaume},
  journal={arXiv preprint arXiv:2509.21753},
  year={2025}
}

@article{sciortino2025active,
  title={Active membrane deformations of a minimal synthetic cell},
  author={Sciortino, Alfredo and Faizi, Hammad A and Fedosov, Dmitry A and Frechette, Layne and Vlahovska, Petia M and Gompper, Gerhard and Bausch, Andreas R},
  journal={Nature Physics},
  pages={1--9},
  year={2025},
  publisher={Nature Publishing Group UK London}
}

@article{grober2023unconventional,
  title={Unconventional colloidal aggregation in chiral bacterial baths},
  author={Grober, Daniel and Palaia, Ivan and U{\c{c}}ar, Mehmet Can and Hannezo, Edouard and {\v{S}}ari{\'c}, An{\dj}ela and Palacci, J{\'e}r{\'e}mie},
  journal={Nature Physics},
  volume={19},
  number={11},
  pages={1680--1688},
  year={2023},
  publisher={Nature Publishing Group UK London}
}

@article{wu2000particle,
  title={Particle diffusion in a quasi-two-dimensional bacterial bath},
  author={Wu, Xiao-Lun and Libchaber, Albert},
  journal={Physical Review Letters},
  volume={84},
  number={13},
  pages={3017},
  year={2000},
  publisher={APS}
}

@article{berezney2024controlling,
  title={Controlling assembly and oscillations of elastic membranes with an active fluid},
  author={Berezney, John and Ray, Sattvic and Kolvin, Itamar and Bowick, Mark and Fraden, Seth and Dogic, Zvonimir},
  journal={arXiv preprint arXiv:2408.14699},
  year={2024}
}

@article{dayel1999diffusion,
  title={Diffusion of green fluorescent protein in the aqueous-phase lumen of endoplasmic reticulum},
  author={Dayel, Mark J and Hom, Erik FY and Verkman, Alan S},
  journal={Biophysical journal},
  volume={76},
  number={5},
  pages={2843--2851},
  year={1999},
  publisher={Elsevier}
}

@article{liu2006quantification,
  title={Quantification of the roles of trabecular microarchitecture and trabecular type in determining the elastic modulus of human trabecular bone},
  author={Liu, Xiaowei S and Sajda, Paul and Saha, Punam K and Wehrli, Felix W and Guo, X Edward},
  journal={Journal of Bone and Mineral Research},
  volume={21},
  number={10},
  pages={1608--1617},
  year={2006},
  publisher={Wiley Online Library}
}

@article{herisson2018direct,
  title={Direct vascular channels connect skull bone marrow and the brain surface enabling myeloid cell migration},
  author={Herisson, Fanny and Frodermann, Vanessa and Courties, Gabriel and Rohde, David and Sun, Yuan and Vandoorne, Katrien and Wojtkiewicz, Gregory R and Masson, Gustavo Santos and Vinegoni, Claudio and Kim, Jiwon and others},
  journal={Nature neuroscience},
  volume={21},
  number={9},
  pages={1209--1217},
  year={2018},
  publisher={Nature Publishing Group US New York}
}

@article{thomas1986ordered,
  title={Ordered bicontinuous double-diamond structure of star block copolymers: a new equilibrium microdomain morphology},
  author={Thomas, Edwin L and Alward, David B and Kinning, David J and Martin, David C and Handlin Jr, Dale L and Fetters, Lewis J},
  journal={Macromolecules},
  volume={19},
  number={8},
  pages={2197--2202},
  year={1986},
  publisher={ACS Publications}
}

@article{werner2018block,
  title={Block copolymer derived 3-D interpenetrating multifunctional gyroidal nanohybrids for electrical energy storage},
  author={Werner, JG and Rodr{\'\i}guez-Calero, GG and Abru{\~n}a, HD and Wiesner, U},
  journal={Energy \& Environmental Science},
  volume={11},
  number={5},
  pages={1261--1270},
  year={2018},
  publisher={Royal Society of Chemistry}
}

@article{siggia1979late,
  title={Late stages of spinodal decomposition in binary mixtures},
  author={Siggia, Eric D},
  journal={Physical review A},
  volume={20},
  number={2},
  pages={595},
  year={1979},
  publisher={APS}
}

@article{zhou2014living,
  title={Living liquid crystals},
  author={Zhou, Shuang and Sokolov, Andrey and Lavrentovich, Oleg D and Aranson, Igor S},
  journal={Biophysical Journal},
  volume={106},
  number={2},
  pages={420a},
  year={2014},
  publisher={Elsevier}
}

@article{wensink2012meso,
  title={Meso-scale turbulence in living fluids},
  author={Wensink, Henricus H and Dunkel, J{\"o}rn and Heidenreich, Sebastian and Drescher, Knut and Goldstein, Raymond E and L{\"o}wen, Hartmut and Yeomans, Julia M},
  journal={Proceedings of the national academy of sciences},
  volume={109},
  number={36},
  pages={14308--14313},
  year={2012},
  publisher={National Academy of Sciences}
}

@article{saintillan2008instabilities,
  title={Instabilities and Pattern Formation in Active Particle Suspensions: Kinetic Theory and Continuum Simulations},
  author={Saintillan, David and Shelley, Michael J},
  journal={Physical Review Letters},
  volume={100},
  number={17},
  pages={178103},
  year={2008},
  publisher={APS}
}

@article{aditi2002hydrodynamic,
  title={Hydrodynamic fluctuations and instabilities in ordered suspensions of self-propelled particles},
  author={Aditi Simha, R and Ramaswamy, Sriram},
  journal={Physical Review Letters},
  volume={89},
  number={5},
  pages={058101},
  year={2002},
  publisher={APS}
}

@article{loren2002determination,
  title={Determination of temperature dependent structure evolution by fast-Fourier transform at late stage spinodal decomposition in bicontinuous biopolymer mixtures},
  author={Lor{\'e}n, Niklas and Langton, Maud and Hermansson, Anne-Marie},
  journal={The Journal of chemical physics},
  volume={116},
  number={23},
  pages={10536--10546},
  year={2002},
  publisher={American Institute of Physics}
}

@article{jinnai2003connectivity,
  title={Connectivity and topology of a phase-separating bicontinuous structure in a polymer mixture: Direct measurements of coordination number, inter-junction distances and Euler characteristic},
  author={Jinnai, Hiroshi and Watashiba, Hideyuki and Kajihara, Takashi and Takahashi, Masaoki},
  journal={The Journal of chemical physics},
  volume={119},
  number={14},
  pages={7554--7559},
  year={2003},
  publisher={American Institute of Physics}
}

@article{chan1996polymerization,
  title={Polymerization-induced phase separation. 1. Droplet size selection mechanism},
  author={Chan, Philip K and Rey, Alejandro D},
  journal={Macromolecules},
  volume={29},
  number={27},
  pages={8934--8941},
  year={1996},
  publisher={ACS Publications}
}

@article{lee2022elastomeric,
  title={Elastomeric electrolytes for high-energy solid-state lithium batteries},
  author={Lee, Michael J and Han, Junghun and Lee, Kyungbin and Lee, Young Jun and Kim, Byoung Gak and Jung, Kyu-Nam and Kim, Bumjoon J and Lee, Seung Woo},
  journal={Nature},
  volume={601},
  number={7892},
  pages={217--222},
  year={2022},
  publisher={Nature Publishing Group UK London}
}

@article{muller2025cell,
  title={Cell-guiding microporous hydrogels by photopolymerization-induced phase separation},
  author={M{\"u}ller, Monica Z and Bernero, Margherita and Xie, Chang and Qiu, Wanwan and Oggianu, Esteban and Rabut, Lucie and Michaels, Thomas CT and Style, Robert W and M{\"u}ller, Ralph and Qin, Xiao-Hua},
  journal={Nature Communications},
  volume={16},
  number={1},
  pages={4923},
  year={2025},
  publisher={Nature Publishing Group UK London}
}

@article{shi2021scaling,
  title={Scaling behavior of stiffness and strength of hierarchical network nanomaterials},
  author={Shi, Shan and Li, Yong and Ngo-Dinh, Bao-Nam and Markmann, J{\"u}rgen and Weissm{\"u}ller, J{\"o}rg},
  journal={Science},
  volume={371},
  number={6533},
  pages={1026--1033},
  year={2021},
  publisher={American Association for the Advancement of Science}
}

@article{stratford2005colloidal,
  title={Colloidal jamming at interfaces: A route to fluid-bicontinuous gels},
  author={Stratford, Kevin and Adhikari, Ronojoy and Pagonabarraga, Ignacio and Desplat, J-C and Cates, Michael E},
  journal={Science},
  volume={309},
  number={5744},
  pages={2198--2201},
  year={2005},
  publisher={American Association for the Advancement of Science}
}

@article{herzig2007bicontinuous,
  title={Bicontinuous emulsions stabilized solely by colloidal particles},
  author={Herzig, Eva M and White, KA and Schofield, Andrew B and Poon, Wilson CK and Clegg, Paul S},
  journal={Nature materials},
  volume={6},
  number={12},
  pages={966--971},
  year={2007},
  publisher={Nature Publishing Group UK London}
}

@article{huang2017bicontinuous,
  title={Bicontinuous structured liquids with sub-micrometre domains using nanoparticle surfactants},
  author={Huang, Caili and Forth, Joe and Wang, Weiyu and Hong, Kunlun and Smith, Gregory S and Helms, Brett A and Russell, Thomas P},
  journal={Nature nanotechnology},
  volume={12},
  number={11},
  pages={1060--1063},
  year={2017},
  publisher={Nature Publishing Group UK London}
}

@article{fernandez2024elastic,
  title={Elastic microphase separation produces robust bicontinuous materials},
  author={Fern{\'a}ndez-Rico, Carla and Schreiber, Sanjay and Oudich, Hamza and Lorenz, Charlotta and Sicher, Alba and Sai, Tianqi and Bauernfeind, Viola and Heyden, Stefanie and Carrara, Pietro and Lorenzis, Laura De and others},
  journal={Nature Materials},
  volume={23},
  number={1},
  pages={124--130},
  year={2024},
  publisher={Nature Publishing Group UK London}
}

@article{royall2018vitrification,
  title={Vitrification and gelation in sticky spheres},
  author={Royall, C Patrick and Williams, Stephen R and Tanaka, Hajime},
  journal={The Journal of chemical physics},
  volume={148},
  number={4},
  year={2018},
  publisher={AIP Publishing}
}

@article{Bray2002,
author = {A. J. Bray},
title = {Theory of phase-ordering kinetics},
journal = {Advances in Physics},
volume = {51},
number = {2},
pages = {481-587},
year = {2002},
publisher = {Taylor & Francis},
}

@article{cahn1958free,
  title={Free energy of a nonuniform system. I. Interfacial free energy},
  author={Cahn, John W and Hilliard, John E},
  journal={The Journal of chemical physics},
  volume={28},
  number={2},
  pages={258--267},
  year={1958},
  publisher={American Institute of Physics}
}

@article{kwon2007coarsening,
  title={Coarsening of bicontinuous structures via nonconserved and conserved dynamics},
  author={Kwon, Yongwoo and Thornton, Katsuyo and Voorhees, Peter W},
  journal={Physical Review E},
  volume={75},
  number={2},
  pages={021120},
  year={2007},
  publisher={APS}
}

@article{araki2001three,
  title={Three-dimensional numerical simulations of viscoelastic phase separation: Morphological characteristics},
  author={Araki, Takeaki and Tanaka, Hajime},
  journal={Macromolecules},
  volume={34},
  number={6},
  pages={1953--1963},
  year={2001},
  publisher={ACS Publications}
}

@article{perrot1999morphology,
  title={Morphology transition observed in a phase separating fluid},
  author={Perrot, F and Beysens, D and Garrabos, Y and Fr{\"o}hlich, T and Guenoun, P and Bonetti, M and Bravais, P},
  journal={Physical Review E},
  volume={59},
  number={3},
  pages={3079},
  year={1999},
  publisher={APS}
}

@article{beysens1997kinetics,
  title={Kinetics and morphology of phase separation in fluids: The role of droplet coalescence},
  author={Beysens, Daniel A},
  journal={Physica A: Statistical Mechanics and its Applications},
  volume={239},
  number={1-3},
  pages={329--339},
  year={1997},
  publisher={Elsevier}
}

@article{tateno2021power,
  title={Power-law coarsening in network-forming phase separation governed by mechanical relaxation},
  author={Tateno, Michio and Tanaka, Hajime},
  journal={Nature communications},
  volume={12},
  number={1},
  pages={912},
  year={2021},
  publisher={Nature Publishing Group UK London}
}

@article{sanchez2012spontaneous,
  title={Spontaneous motion in hierarchically assembled active matter},
  author={Sanchez, Tim and Chen, Daniel TN and DeCamp, Stephen J and Heymann, Michael and Dogic, Zvonimir},
  journal={Nature},
  volume={491},
  number={7424},
  pages={431--434},
  year={2012},
  publisher={Nature Publishing Group UK London}
}

@article{adkins2022dynamics,
  title={Dynamics of active liquid interfaces},
  author={Adkins, Raymond and Kolvin, Itamar and You, Zhihong and Witthaus, Sven and Marchetti, M Cristina and Dogic, Zvonimir},
  journal={Science},
  volume={377},
  number={6607},
  pages={768--772},
  year={2022},
  publisher={American Association for the Advancement of Science}
}

@article{zhao2024asymmetric,
  title={Asymmetric fluctuations and self-folding of active interfaces},
  author={Zhao, Liang and Gulati, Paarth and Caballero, Fernando and Kolvin, Itamar and Adkins, Raymond and Marchetti, M Cristina and Dogic, Zvonimir},
  journal={Proceedings of the National Academy of Sciences},
  volume={121},
  number={51},
  pages={e2410345121},
  year={2024},
  publisher={National Academy of Sciences}
}

@article{sessa2026interfacial,
  title = {Interfacial instability of confined three-dimensional active droplets},
  author = {Sessa, Bennett C. and Cao, Federico and Pelcovits, Robert A. and Powers, Thomas R. and Duclos, Guillaume},
  journal = {Physical Review Research},
  volume = {8},
  issue = {1},
  pages = {013120},
  numpages = {19},
  year = {2026},
  month = {Feb},
  publisher = {American Physical Society}
}

@article{kempf2019active,
  title={Active matter invasion},
  author={Kempf, Felix and Mueller, Romain and Frey, Erwin and Yeomans, Julia M and Doostmohammadi, Amin},
  journal={Soft matter},
  volume={15},
  number={38},
  pages={7538--7546},
  year={2019},
  publisher={Royal Society of Chemistry}
}

@article{zhang2011three,
  title={Three-dimensional bicontinuous ultrafast-charge and-discharge bulk battery electrodes},
  author={Zhang, Huigang and Yu, Xindi and Braun, Paul V},
  journal={Nature nanotechnology},
  volume={6},
  number={5},
  pages={277--281},
  year={2011},
  publisher={Nature Publishing Group UK London}
}

@article{jinnai1995direct,
  title={Direct observation of three-dimensional bicontinuous structure developed via spinodal decomposition},
  author={Jinnai, Hiroshi and Nishikawa, Yukihiro and Koga, Tsuyoshi and Hashimoto, Takeji},
  journal={Macromolecules},
  volume={28},
  number={13},
  pages={4782--4784},
  year={1995},
  publisher={ACS Publications}
}

@article{hajduk1994gyroid,
  title={The gyroid: a new equilibrium morphology in weakly segregated diblock copolymers},
  author={Hajduk, Damian A and Harper, Paul E and Gruner, Sol M and Honeker, Christian C and Kim, Gia and Thomas, Edwin L and Fetters, Lewis J},
  journal={Macromolecules},
  volume={27},
  number={15},
  pages={4063--4075},
  year={1994},
  publisher={ACS Publications}
}

@article{sai2013hierarchical,
  title={Hierarchical porous polymer scaffolds from block copolymers},
  author={Sai, Hiroaki and Tan, Kwan Wee and Hur, Kahyun and Asenath-Smith, Emily and Hovden, Robert and Jiang, Yi and Riccio, Mark and Muller, David A and Elser, Veit and Estroff, Lara A and others},
  journal={Science},
  volume={341},
  number={6145},
  pages={530--534},
  year={2013},
  publisher={American Association for the Advancement of Science}
}

@article{liu2025nonequilibrium,
  title={Nonequilibrium dynamics of membraneless active droplets},
  author={Liu, Chenxi and Cao, Ding and Liu, Siyu and Wu, Yilin},
  journal={arXiv preprint arXiv:2511.04181},
  year={2025}
}

@incollection{tayar2022assembling,
  title={Assembling microtubule-based active matter},
  author={Tayar, Alexandra M and Lemma, Linnea M and Dogic, Zvonimir},
  booktitle={Microtubules: Methods and Protocols},
  pages={151--183},
  year={2022},
  publisher={Springer}
}

@article{liu2012concentration,
  title={Concentration dependence of the interfacial tension for aqueous two-phase polymer solutions of dextran and polyethylene glycol},
  author={Liu, Yonggang and Lipowsky, Reinhard and Dimova, Rumiana},
  journal={Langmuir},
  volume={28},
  number={8},
  pages={3831--3839},
  year={2012},
  publisher={ACS Publications}
}

@article{gidi2018efficient,
  title={Efficient one-step PEG-silane passivation of glass surfaces for single-molecule fluorescence studies},
  author={Gidi, Yasser and Bayram, Serene and Ablenas, Christopher J and Blum, Amy Szuchmacher and Cosa, Gonzalo},
  journal={ACS applied materials \& interfaces},
  volume={10},
  number={46},
  pages={39505--39511},
  year={2018},
  publisher={ACS Publications}
}

@article{singh2019hydrodynamically,
  title={Hydrodynamically interrupted droplet growth in scalar active matter},
  author={Singh, Rajesh and Cates, ME},
  journal={Physical Review Letters},
  volume={123},
  number={14},
  pages={148005},
  year={2019},
  publisher={APS}
}

@article{tiribocchi2015active,
  title = {Active Model H: Scalar Active Matter in a Momentum-Conserving Fluid},
  author = {Tiribocchi, Adriano and Wittkowski, Raphael and Marenduzzo, Davide and Cates, Michael E.},
  journal = {Physical Review Letters},
  volume = {115},
  issue = {18},
  pages = {188302},
  numpages = {5},
  year = {2015},
  month = {Oct},
  publisher = {American Physical Society}
}

@article{berg2019ilastik,
  title={Ilastik: interactive machine learning for (bio) image analysis},
  author={Berg, Stuart and Kutra, Dominik and Kroeger, Thorben and Straehle, Christoph N and Kausler, Bernhard X and Haubold, Carsten and Schiegg, Martin and Ales, Janez and Beier, Thorsten and Rudy, Markus and others},
  journal={Nature methods},
  volume={16},
  number={12},
  pages={1226--1232},
  year={2019},
  publisher={Nature Publishing Group US New York}
}

@inproceedings{cignoni2008meshlab,
  title={Meshlab: an open-source mesh processing tool.},
  author={Cignoni, Paolo and Callieri, Marco and Corsini, Massimiliano and Dellepiane, Matteo and Ganovelli, Fabio and Ranzuglia, Guido and others},
  booktitle={Eurographics Italian chapter conference},
  volume={2008},
  pages={129--136},
  year={2008},
  organization={Salerno}
}

@software{Imaris,
    title = {Imaris},
    author = {{Oxford Instruments}},
    version = {10.2},
    year = {2024},
    publisher = {Oxford Instruments},
    url = {https://imaris.oxinst.com}
}

@misc{gut_matlab,
author = {Mitchell, Noah P and Cislo, Dillon},
title = {gut\_matlab. {G}itHub repository},
note = { https://github.com/npmitchell/gut\_matlab (accessed on 6/6/2025)},
year = {2025},
}

@inproceedings{rusinkiewicz2004estimating,
  title={Estimating curvatures and their derivatives on triangle meshes},
  author={Rusinkiewicz, Szymon},
  booktitle={Proceedings. 2nd International Symposium on 3D Data Processing, Visualization and Transmission, 2004. 3DPVT 2004.},
  pages={486--493},
  year={2004},
  organization={IEEE}
}

@article{gulati2025active,
  title={Active fluids form system-spanning filamentary networks},
  author={Gulati, Paarth and Caballero, Fernando and Cristina Marchetti, M},
  journal={Physical Review Letters},
  volume={134},
  number={13},
  pages={138301},
  year={2025},
  publisher={APS}
}

@article{putzig2016instabilities,
  title={Instabilities, defects, and defect ordering in an overdamped active nematic},
  author={Putzig, Elias and Redner, Gabriel S and Baskaran, Arvind and Baskaran, Aparna},
  journal={Soft matter},
  volume={12},
  number={17},
  pages={3854--3859},
  year={2016},
  publisher={Royal Society of Chemistry}
}

@article{santhosh2020activity,
  title={Activity induced nematic order in isotropic liquid crystals},
  author={Santhosh, Sreejith and Nejad, Mehrana R and Doostmohammadi, Amin and Yeomans, Julia M and Thampi, Sumesh P},
  journal={Journal of Statistical Physics},
  volume={180},
  number={1},
  pages={699--709},
  year={2020},
  publisher={Springer}
}

@article{vafa2021fluctuations,
  title={Fluctuations can induce local nematic order and extensile stress in monolayers of motile cells},
  author={Vafa, Farzan and Bowick, Mark J and Shraiman, Boris I and Marchetti, M Cristina},
  journal={Soft Matter},
  volume={17},
  number={11},
  pages={3068--3073},
  year={2021},
  publisher={Royal Society of Chemistry}
}

@article{srivastava2016negative,
  title={Negative stiffness and modulated states in active nematics},
  author={Srivastava, Pragya and Mishra, Prashant and Marchetti, M Cristina},
  journal={Soft matter},
  volume={12},
  number={39},
  pages={8214--8225},
  year={2016},
  publisher={Royal Society of Chemistry}
}

@article{caballero2022activity,
  title={Activity-suppressed phase separation},
  author={Caballero, Fernando and Marchetti, M Cristina},
  journal={Physical Review Letters},
  volume={129},
  number={26},
  pages={268002},
  year={2022},
  publisher={APS}
}

@article{blow2014biphasic,
  title={Biphasic, lyotropic, active nematics},
  author={Blow, Matthew L and Thampi, Sumesh P and Yeomans, Julia M},
  journal={Physical Review Letters},
  volume={113},
  number={24},
  pages={248303},
  year={2014},
  publisher={APS}
}

@article{coelho2023active,
  title={Active nematics on flat surfaces: from droplet motility and scission to active wetting},
  author={Coelho, Rodrigo CV and Figueiredo, H{\'e}lio RJC and Telo da Gama, Margarida M},
  journal={Physical Review Research},
  volume={5},
  number={3},
  pages={033165},
  year={2023},
  publisher={APS}
}

@article{caballero2024cupss,
  title={cuPSS: a package for pseudo-spectral integration of stochastic PDEs},
  author={Caballero, Fernando},
  journal={Benchmarking},
  volume={11},
  pages={14},
  year={2024}
}

@article{tayar2023controlling,
  title={Controlling liquid--liquid phase behaviour with an active fluid},
  author={Tayar, Alexandra M and Caballero, Fernando and Anderberg, Trevor and Saleh, Omar A and Cristina Marchetti, M and Dogic, Zvonimir},
  journal={Nature Materials},
  volume={22},
  number={11},
  pages={1401--1408},
  year={2023},
  publisher={Nature Publishing Group UK London}
}

@article{gulati2024traveling,
  title={Traveling waves at the surface of active liquid crystals},
  author={Gulati, Paarth and Caballero, Fernando and Kolvin, Itamar and You, Zhihong and Marchetti, M Cristina},
  journal={Soft Matter},
  volume={20},
  number={38},
  pages={7703--7714},
  year={2024},
  publisher={Royal Society of Chemistry}
}

@article{cates2025active,
  title={Active phase separation: new phenomenology from non-equilibrium physics},
  author={Cates, Michael Elmhurst and Nardini, Cesare},
  journal={Reports on Progress in Physics},
  volume={88},
  number={5},
  pages={056601},
  year={2025},
  publisher={IOP Publishing}
}

@article{giomi2012banding,
  title={Banding, excitability and chaos in active nematic suspensions},
  author={Giomi, Luca and Mahadevan, L and Chakraborty, Bulbul and Hagan, MF},
  journal={Nonlinearity},
  volume={25},
  number={8},
  pages={2245},
  year={2012},
  publisher={IOP Publishing}
}

@misc{dryad,
author = {Zhao, Liang},
title = {Bicontinuity in active phase separation},
note = {Dryad. https://doi.org/10.5061/dryad.612jm64kw},
year = {2026},
}

@software{Gulati_3D_cupss,
title = {{3D\_cuPSS}. {G}itHub repository},
author = {Gulati, Paarth},
license = {MIT},
year = {2026},
note = {\url{https://github.com/paarthgulati/bicontinuity_3D_Active_PS}}
}
\bibliographystyle{sciencemag}

\paragraph*{Funding:}
This work was primarily supported by the US Department of Energy, Office of Basic Energy Sciences under award number DE-SC0019733. Development and optimization of the two-phase system of active liquid–liquid phase separation was supported by NSF-ISS-2224350. OS and MT acknowledge support of NSF-MRSEC DMR 2308708. PG was
funded, in part, by the Tarbutton Interdisciplinary Postdoctoral Fellowship at Emory College of Arts and Sciences. We also acknowledge the use of the biosynthesis facility supported by NSF-MRSEC-2011846 and NRI-MCDB Microscopy Facility supported by the NIH Shared Instrumentation Grant 1S10OD019969-01A1. Use was made of computational facilities purchased with funds from the NSF(CNS-1725797) and administered by the Center for Scientific Computing (CSC). The CSC is supported by the California NanoSystems Institute and the Materials Research Science and Engineering Center (NSF-DMR 2308708).

\paragraph*{Author contributions:}
P.G., L.Z., Z.D., and M.C.M designed research; P.G., and L.Z. performed research; P.G., L.Z., and M.T. analyzed data; P.G., L.Z., M.T., O.A.S, Z.D., and M.C.M. wrote the paper.

\paragraph*{Competing interests:}
The authors declare that they have no competing interests.
\paragraph*{Data and materials availability:}
Experimental data of all bicontinuous structures and corresponding analysis codes are deposited in Dryad (DOI: 10.5061/dryad.612jm64kw) \cite{dryad}. The codes for the numerical simulations are available at~\cite{Gulati_3D_cupss}. All other data needed to evaluate and reproduce the results in the paper are present in the paper and/or supporting information. 
% Specify where the data, software, physical samples, simulation outputs or other materials
% underlying the paper are archived.
% They must be publicly accessible when the paper is published (without embargo) and enable
% readers to reproduce all the results in the paper.
% Contact the editor if you’re unsure what needs to be shared.

% Our preference is for digital material to be deposited in a suitable non-profit online data or software repository that issues the material with a DOI. Alternatively, an institutional repository, subject-based archive, commercial repository etc. is acceptable, as are (short) supplementary tables or a machine-readable supplementary data file.
% ‘Available on request’ or personal web pages are not allowed.

% Cite the relevant DOI \cite{dataset}, URL \cite{example_url} or reference \cite{example2} in this statement.
% These \textit{do not} count towards the reference limit if they are only cited in the acknowledgements.
% Be specific and state a unique identifier -- such as an accession number, software version number
% or observation ID -- so readers can easily retrieve the exact material used.

\subsection*{Supplementary materials}
Supplementary Text\\
Figs. S1 to S6\\
% References \textit{(7-\arabic{enumiv})}\\  
Movies S1 to S2\\

\clearpage

%%%%%%%%%%%%%%%% START OF SUPPLEMENT %%%%%%%%%%%%%%%

% Figures, tables, equations and pages in the supplement are numbered S1, S2 etc.
\renewcommand{\thefigure}{S\arabic{figure}}
\renewcommand{\thetable}{S\arabic{table}}
\renewcommand{\theequation}{S\arabic{equation}}
\renewcommand{\thepage}{S\arabic{page}}
\setcounter{figure}{0}
\setcounter{table}{0}
\setcounter{equation}{0}
\setcounter{page}{1} 
%%%%%%%%%%%%%%%% SUPPLEMENT TITLE PAGE %%%%%%%%%%%%%%%

\begin{center}
\section*{Supplementary Materials for\\ \scititle}

Paarth Gulati$^{1\dagger}$,
Liang Zhao$^{1\dagger}$,
Michio Tateno$^{2}$,
Omar A. Saleh$^{1,2,3}$,\\
Zvonimir Dogic$^{1,4}$,
M. Cristina Marchetti$^{1,4\ast}$
\small$^\ast$Corresponding author: mcmarche@physics.ucsb.edu\\
\small$^\dagger$These authors contributed equally to this work.
\end{center}

\subsubsection*{This PDF file includes:}
Supplementary Text\\
Figs. S1 to S6\\
Captions for Movies S1 to S2\\
\subsubsection*{Other Supplementary Materials for this manuscript:}
Movies S1 to S2\\

\newpage

\subsection*{Supplementary Text}

\subsubsection*{Capillary length measurement}
\begin{figure*} 
	\centering
	\includegraphics[width=0.6\textwidth]{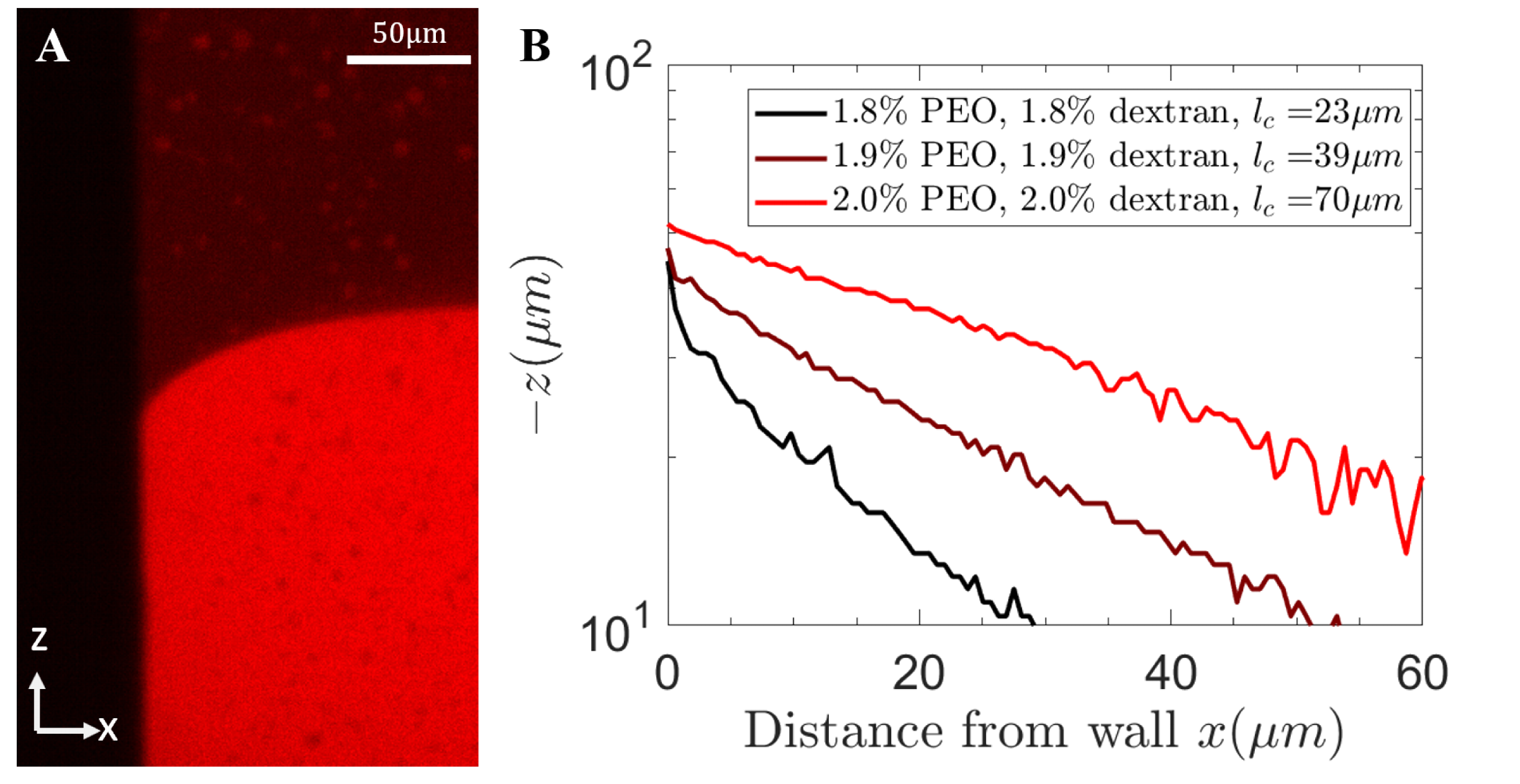} 
	\caption{Measurement of capillary lengths between the two separated phases. (A) A two-dimensional crosssection of the interface profile between two bulk separated phases near the wall in a static sample containing $1.9$\% PEO and $1.9$\% dextran. The
dextran phase de-wets the FEP wall. (B) The interface height z versus distance from the wall x plotted on a log-linear scale, demonstrating the exponential dependence of
the height near the wall. The capillary lengths of the samples with three different polymer concentrations are $23~\mu$m, $39~\mu$m, and $70~\mu$m each. The values are obtained by fitting an
exponential function to the interface profile near the wall. The measured samples contain no MTs or KSA.}
	\label{fig:sup_capillary} % 
\end{figure*}

 The capillary lengths $l_c$ between the PEO and dextran phase were measured using a previously published method \cite{zhao2024asymmetric}. Passive samples containing no MT or KSA and containing $1.8$\%-$2.0$\% PEO and $1.8$\%-$2.0$\% dextran were placed in transparent fluorinated ethylene propylene (FEP) tubes with inner diameters $2.4$~mm. The samples were centrifuged at $2000$~g relative centrifugal force for $1$ min (Fisher Scientific, 05-090-128) to bulk separate the PEO-rich and dextran-rich phases. The samples were then imaged on a laser sheet microscope (Zeiss Z.1 Lightsheet). The dextran phases de-wetted the FEP wall (Fig.~\ref{fig:sup_capillary} A). The interface profiles near the wall were then fitted by exponential functions to obtain $l_c$ (Fig.~\ref{fig:sup_capillary} B).

\subsubsection*{Gravity effect on bicontinuous structure}
\begin{figure*} 
	\centering
\includegraphics[width=0.9\textwidth]{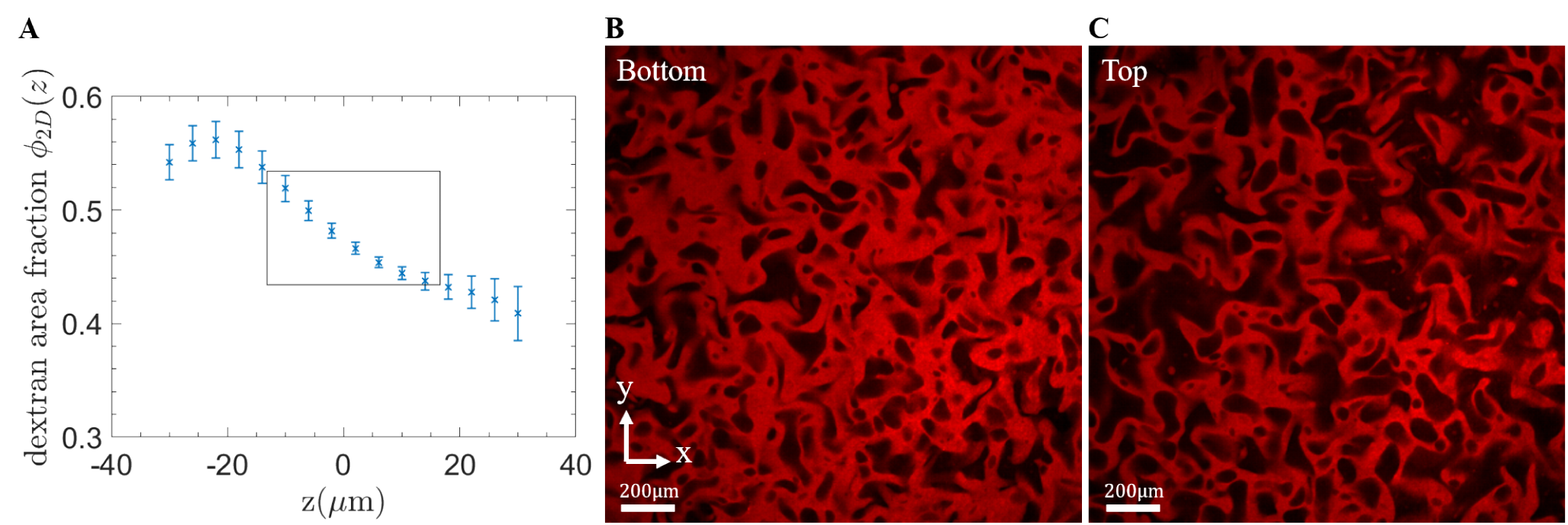} 
	\caption{\textbf{Gravity causes inhomogeneous dextran phase fraction at different $z$ positions.} (A) Dextran area fraction $\phi_{2D}$ vs. $z$, averaged over 5 minutes. $z=0$ is the middle of the sample. The gravitational field points along negative $z$. The volume fraction of dextran phase $\phi_a=0.48$. The black box indicates where $\phi_a-0.05<\phi_{2D} (z)<\phi_a+0.05$. (B, C) 2D horizontal cross-sections of the sample in (A) at $t=2.5$~min and $z$ position where $\phi_{2D} (z)=\phi_a+0.05$ (B) and $\phi_{2D} (z)=\phi_a-0.05$ (C). The sample has an initial concentration of $42$~nM KSA, $1.8$\% PEO and $1.8$ \% dextran.}
	\label{fig:gravity} % 
\end{figure*}

In the experiment, since the dextran phase has a slightly larger density than the PEO phase, it gradually sediments in the $-z$ direction due to gravity over time. This causes an inhomogeneous amount of dextran phase at different $z$ positions (Fig.~\ref{fig:gravity} B, C). To minimize this effect, we define the area fraction of the dextran phase at each $z$ as
\begin{equation}
	\phi_{2D} (z) = A_d (z)/A_{total} (z)
	\label{eq:fd2d} 
\end{equation}
where $A_d(z)$ is the area of dextran phase and $A_{total}(z)$ is the area of the whole sample. 
In each sample, we pick the $z$ position where $\phi_a-0.05<\phi_{2D} (z)<\phi_a+0.05$ (Fig.~\ref{fig:gravity}
 A) for all analysis. 
 
\begin{figure*} 
	\centering
\includegraphics[width=0.9\textwidth]{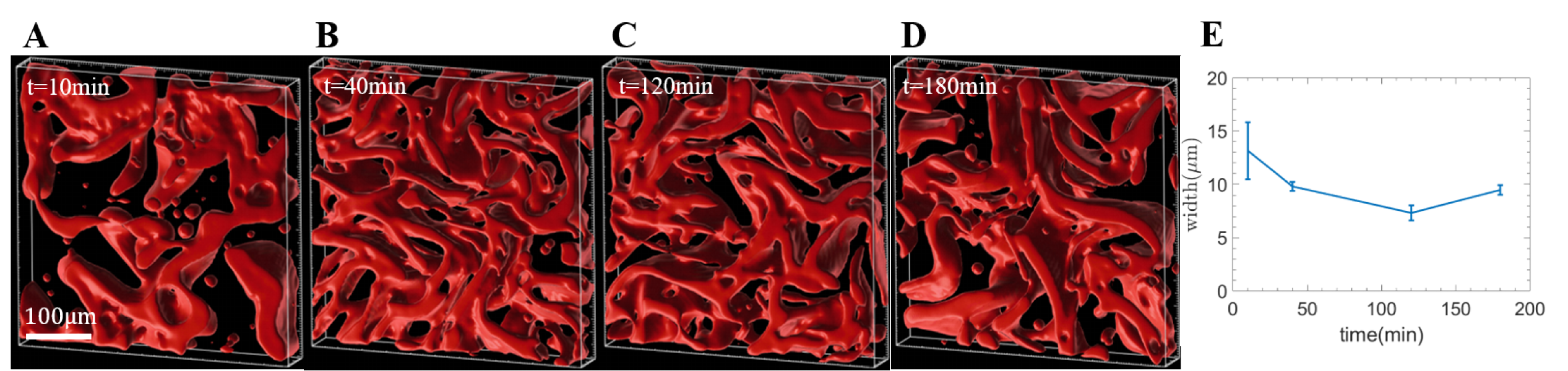} 
	\caption{\textbf{Bicontinuous structure reached steady state.} 3D visualizations of the dextran phase (red) at 4 different time points (A-D). The network structure reached a steady state at $t=40$~min and remained similar over 2 hours (B-D) together with the active width $w_a$ (E). The sample initially contain 92~nM KSA, 1.8\% PEO and 1.8\% dextran and is recombined to obtain active fraction $\phi_a=0.28$. 
		}
	\label{fig:sup_evolution} % 
\end{figure*}

To verify that the bicontinuous structure persists in steady state, we image the sample at $t$=10~min, 40~min, 120~min and 180~min (Fig.~\ref{fig:sup_evolution}). In each acquisition, we image the sample for 5~min, and we keep the sample on a rotor at 24~rpm between the acquisitions. We observe the bicontinuity persists and $w_a$ remains the same over 2 hours (Fig.~\ref{fig:sup_evolution}), which reveals that the sample is in steady state in absence of sedimentation, consistent with simulation results (Fig.~\ref{setup} B). 
 
\begin{figure*} % Do not use \begin{figure*}
	\centering
	\includegraphics[width=0.6\textwidth]{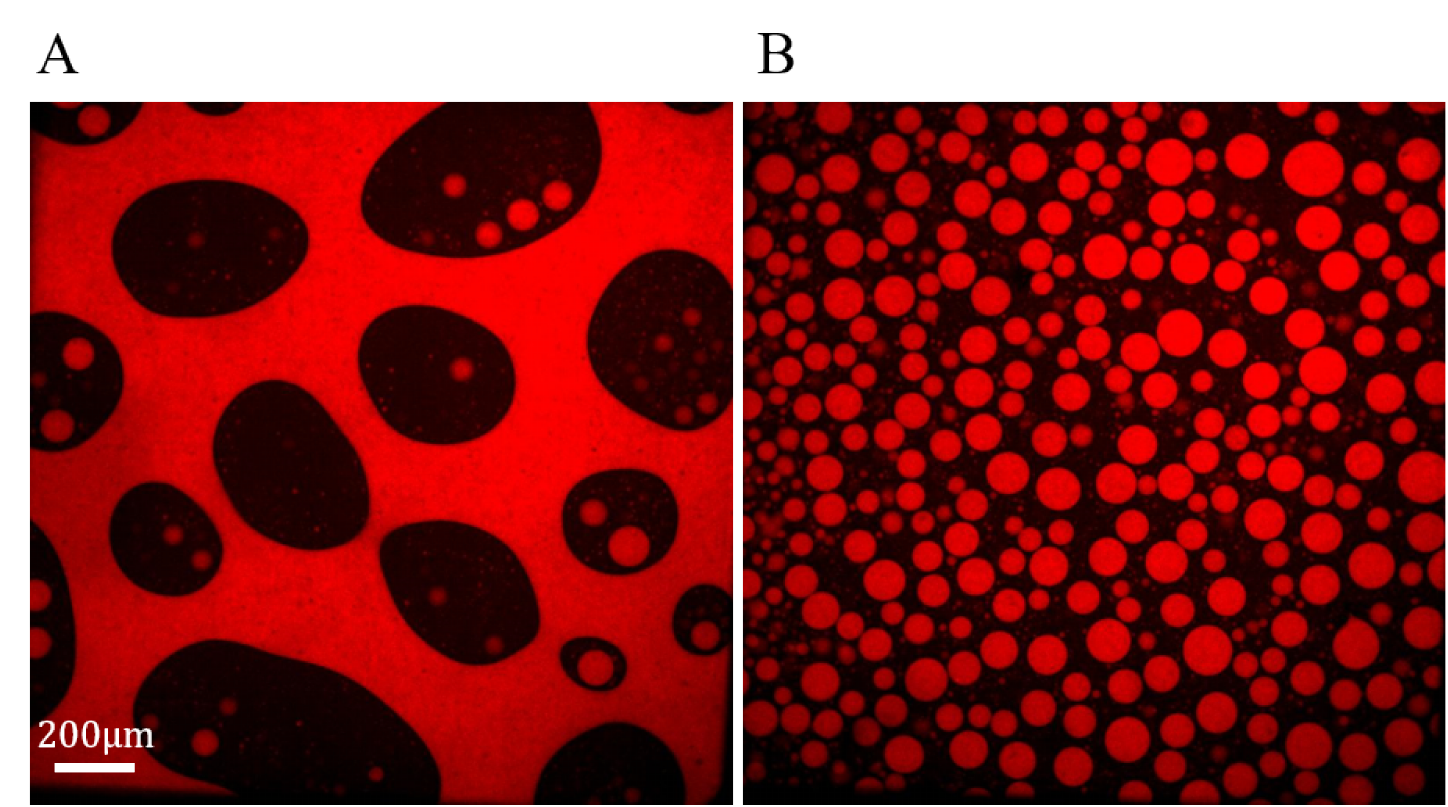} 
	\caption{\textbf{Passive phase separated samples form droplets} (A,B) Two dimensional crosssections of passive samples with initially $1.8$\% PEO and $1.8$\% dextran, no KSA. The passive phase forms droplets when $\phi_a=0.56>0.5$ in (A) and the active phase form droplets when $\phi_a=0.41<0.5$ in (B).
		}
	\label{fig:sup_passive} % give each figure a logical label name
\end{figure*}

\begin{figure*}
    \centering
\includegraphics[width=0.9\textwidth]{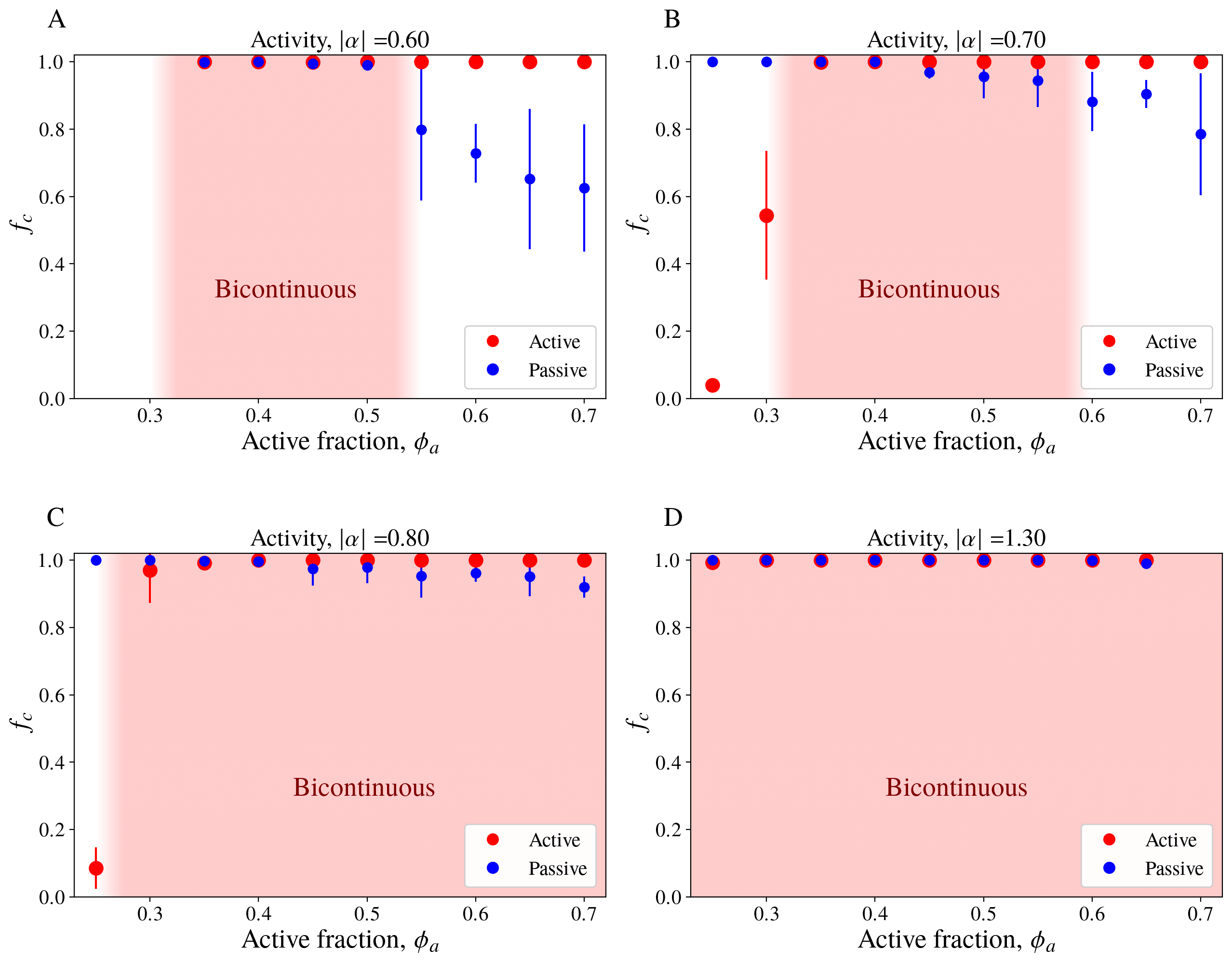} 
	\caption{\textbf{Bicontinuity for increasing activity.} The dependence of active/passive domain connectedness fractions on the activity. For large activities, the bicontinuous state for the entire range of volume fractions explored numerically i.e. the spinodal region of the passively phase separating system. }
	\label{fig:sup_bicontinuity_activities} % 
\end{figure*}

\begin{figure*} 
	\centering
\includegraphics[width=0.9\textwidth]{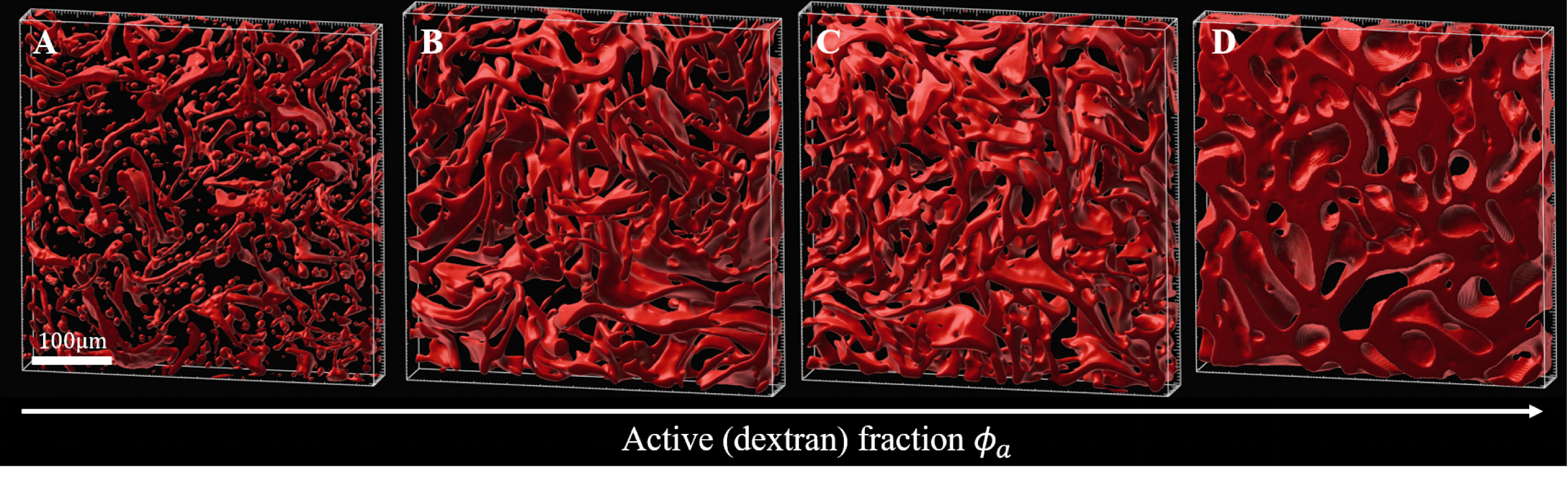} 
	\caption{\textbf{Visualization of active phase (red) morphology over varying $\phi_a$.} Samples contain $\phi_a=$0.16 (A), 0.29 (B), 0.37 (C), 0.60 (D), corresponding to data points in Fig.~\ref{fig:bicontinuous}D. All samples initially contain 183~nM KSA, 1.8\% PEO, and 1.8\% dextran and are recombined to obtain various $\phi_a$.}
	\label{fig:sup_fraction} % 
\end{figure*}

\clearpage % Clear all remaining figures and tables then start a new page

\paragraph{Caption for Movie S1.}
\textbf{Three-dimensional visualization of active phases (red) forming dynamic networks in experiment at different KSA concentrations}. The movie starts at $1.5$~h after all sample were prepared. The samples initially contain $1.8$\% PEO, $1.8$\% dextran, $42-183$~nM KSA. 
\paragraph{Caption for Movie S2.}
\textbf{Three-dimensional visualization of active phase (red) forming a dynamic network in simulation}. The sample is same as Fig.~\ref{setup} A-B. %The sample has activity $|\alpha|=1.30$, surface tension $\gamma=1.0$, and active fraction $\phi_a=0.40$.

\end{document}